\documentclass[11pt]{article}

\usepackage[utf8]{inputenc}
\usepackage[T1]{fontenc}

\ifdefined\TimesFont
  \usepackage{times}
\fi

\usepackage{amsmath,amsthm,amssymb,amsfonts,bm,mathtools}
\usepackage{mathrsfs}

\IfFileExists{dsfont.sty}{%
  \usepackage{dsfont}
}{%
  
}

\usepackage[margin=1in]{geometry}
\usepackage{ragged2e}
\usepackage{enumitem}
\usepackage{xcolor}
\usepackage{xargs}
\usepackage{microtype}
\usepackage{nicefrac}
\usepackage{authblk}
\usepackage{verbatim}
\usepackage{comment}

\ifdefined\ParSkip
  \usepackage{parskip}
\fi

\usepackage{graphicx}
\usepackage{caption}
\usepackage{subcaption}
\usepackage{booktabs}
\usepackage{array}
\usepackage{multirow}
\usepackage{placeins}
\usepackage{float}
\usepackage{appendix}
\usepackage{tikz}

\usepackage{algorithm}
\usepackage{algorithmic}

\usepackage[authoryear,round]{natbib}
\usepackage{url}

\usepackage{xr-hyper}

\usepackage[
  colorlinks=true,
  citecolor=blue,
  urlcolor=blue,
  linkcolor=blue
]{hyperref}

\usepackage{cleveref}

\theoremstyle{plain}
\newtheorem{theorem}{Theorem}
\newtheorem{lemma}{Lemma}
\newtheorem{corollary}{Corollary}

\theoremstyle{definition}
\newtheorem{definition}{Definition}
\newtheorem{procedure}{Procedure}
\newtheorem{remark}{Remark}
\newtheorem{example}{Example}

\crefname{procedure}{procedure}{procedures}
\Crefname{procedure}{Procedure}{Procedures}

\providecommand{\assumptionnumber}{}
\newtheorem*{assumption*}{\assumptionnumber}



\makeatletter
\newcommand*\rel@kern[1]{%
  \kern#1\dimexpr\macc@kerna\relax
}

\newcommand*\widebar[1]{%
  \begingroup
  \def\mathaccent##1##2{%
    \rel@kern{0.8}%
    \overline{%
      \rel@kern{-0.8}%
      \macc@nucleus
      \rel@kern{0.2}%
    }%
    \rel@kern{-0.2}%
  }%
  \macc@depth\@ne
  \let\math@bgroup\@empty
  \let\math@egroup\macc@set@skewchar
  \mathsurround\z@
  \frozen@everymath{\mathgroup\macc@group\relax}%
  \macc@set@skewchar\relax
  \let\mathaccentV\macc@nested@a
  \macc@nested@a\relax111{#1}%
  \endgroup
}
\makeatother

\usepackage{amsthm}
\usepackage{comment}
\usepackage{appendix}
\usepackage{array}

\makeatletter
\@ifundefined{procedure}{%
  \theoremstyle{definition}
  \newtheorem{procedure}{Procedure}
}{}
\makeatother

\title{False Discovery Rate Control in Ordered Sequential Testing:
Rejection-Count and Fallback Procedures}

\author[1]{Anjana Grandhi}
\author[2]{Wenge Guo\thanks{Author e-mail addresses: anjana.grandhi@merck.com, wenge.guo@njit.edu, loh@njit.edu; sanat.sarkar@temple.edu}}
\author[2]{Ji Meng Loh}
\author[3]{Sanat Sarkar}
\affil[1]{BARDS, Merck Research Laboratories\\
        Rahway, NJ 07065, USA}
\affil[2]{Department of Mathematical Sciences\\
       New Jersey Institute of Technology\\
      Newark, NJ 07102, USA}
\affil[3]{Department of Statistics, Operations and Data Science, Temple University, Philadelphia, PA, USA}

\date{\today}

\begin{document}

\maketitle

\begin{abstract}
We study false discovery rate (FDR) control for a fixed
family of hypotheses tested sequentially in a prespecified
order. Each decision uses only the current and preceding
$p$-values, is not subsequently revised, and permits
continued testing after non-rejections. We distinguish
generalized fixed-sequence procedures based on preceding
rejection counts from fallback-type procedures that
additionally use recursively transferred testing levels.
For rejection-count procedures, we study a position-dependent
calibration under independence and a linear rule calibrated
through null marginal distribution functions under positive
left-tail dependence under the null. We clarify their
connections to existing fixed-sequence and online FDR
methods. The fallback construction combines the total
number of preceding discoveries with level transfer along
consecutive rejections, allowing thresholds to reflect
the arrangement of earlier outcomes. We establish its
FDR control under history-based positive regression
dependence and, with a position-specific correction,
under arbitrary dependence. For all four constructions,
the guarantees extend to every deterministic testing time
within the planned horizon. Gaussian simulations assess
the effects of signal placement, dependence, and allocation
weights. They demonstrate power gains over selected
benchmarks while identifying the conservativeness of
the arbitrary-dependence correction.
\end{abstract}


\section{Introduction}
\label{s:intro}

Multiple testing problems often involve hypotheses that must be evaluated in a prespecified order, reflecting scientific priorities, external information, or the sequence in which data become available. Examples include screening candidates based on prior evidence, monitoring a process at successive testing occasions, and investigating a planned sequence of scientific questions. In such settings, decisions may be required before future $p$-values become available, while later hypotheses may remain of interest even if earlier hypotheses are not rejected.

These settings call for procedures that balance the identification of discoveries against the risk of false rejections. Controlling the family-wise error rate (FWER), the probability of at least one false rejection, provides stringent protection but may substantially limit discoveries when many hypotheses are tested. The false discovery rate (FDR), defined as the expected proportion of false discoveries among all rejections, offers a less restrictive alternative when some false discoveries can be tolerated \citep{Benjamini_1995}. Classical procedures, such as the Benjamini--Hochberg (BH) procedure and its Benjamini--Yekutieli (BY) modification, determine rejections using the full collection of $p$-values \citep{Benjamini_1995,Benjamini_2001}. Consequently, they do not directly accommodate irrevocable decisions made before future $p$-values are observed.

Prespecified ordering has a long history in FWER-controlling multiple testing; see \citet{tamhane2018advances} for a review. Conventional fixed-sequence testing stops at the first non-rejection \citep{Maurer_1995}, whereas generalized fixed-sequence procedures relax this restriction and fallback procedures redistribute testing levels following rejections \citep{Wiens_2003,Wiens_2005,qiu2015generalized}. For FDR control, \citet{lynch2017control} develop procedures that stop after a prespecified number of non-rejections, while \citet{Farcomeni_2013} considers a common testing threshold with a stopping boundary determined by cumulative non-rejections. These developments illustrate how rejection thresholds and stopping rules can be used separately to regulate testing along a prespecified sequence.

Other approaches that exploit a supplied ordering include SeqStep, sequential selection rules, and accumulation tests \citep{barber2015controlling,gsell2016sequential,li2017accumulation}, whose power depends on the quality of the ordering \citep{lei2016power}. However, their final rejection cutoffs generally depend on subsequently observed $p$-values. In contrast, the procedures considered here make an irrevocable decision on $H_i$ using only information available through the $i$th testing occasion. The essential distinction is therefore the information available at the time of each decision.

Online FDR procedures share this sequential information constraint. Methods such as LOND, LORD, and SAFFRON allocate testing levels based on preceding information and accommodate potentially unbounded testing sequences \citep{javanmard2015online,javanmard2018online,ramdas2018saffron}. More recent developments address positive and local dependence, e-value-based testing under arbitrary dependence, and improvements through online e-closure \citep{fisher2024online,xu2024online,xu2026improving}. Thus, conditional super-uniformity given the full history of preceding $p$-values is not a universal requirement of online FDR methods. A comprehensive review is provided by \citet{robertson2023online}.

In this paper, we consider a fixed, known testing horizon $m$, which allows rejection thresholds to be calibrated using the number of remaining hypotheses. The finite horizon serves as a calibration device rather than defining a fundamentally different testing framework. Although our main results concern the terminal rejection set, the FDR guarantees also hold at every deterministic time $t\le m$, with thresholds calibrated to the original horizon. The proposed procedures therefore provide pointwise-in-time FDR control within a planned finite sequence, as in online testing, without asserting validity under arbitrary data-dependent stopping or an unknown horizon.

We develop two classes of procedures distinguished by their use of preceding testing outcomes. The first consists of generalized fixed-sequence procedures whose rejection thresholds depend on the testing position and the number of preceding rejections. The second consists of fallback-type procedures that additionally incorporate recursively transferred testing levels, allowing thresholds to depend on the arrangement of preceding rejections. Our focus is on these two structures rather than an exhaustive classification of sequential multiple testing methods.

For generalized fixed-sequence procedures, we introduce a position-dependent calibration under independence and a linear threshold rule under positive left-tail dependence under the null (PLTDN). The former refines the calibration of a generalized fixed-sequence procedure proposed by \citet{lynch2017control}. The latter builds on the dependence framework originally introduced as positive tail dependence under the null (PTDN) by \citet{GhoshSarkar2025}, here referred to as PLTDN to emphasize its left-tail formulation. An important feature of this framework is that it accommodates suitably calibrated $p$-values, rather than restricting the dependence condition to the original $p$-values. Such calibration allows information about the dependence structure to be incorporated into the testing procedure. By combining PLTDN with calibration based on the null distributions of the relevant test statistics, we obtain a procedure that retains the linear threshold structure of LOND while accounting for dependence through the calibrated evidence. This development is also related to the shifted-testing methodology of \citet{Sarkar_2025,GhoshSarkar2025}. Existing LOND calibrations under PRDS and arbitrary dependence serve as benchmarks rather than additional methodological contributions.

For fallback-type procedures, we combine recursive level transfer along consecutive rejections with a multiplier determined by the total number of preceding discoveries. Consequently, testing histories with the same rejection count may produce different thresholds, depending on the arrangement of rejections. We establish FDR control under history-based PRDS and, with a position-specific correction, under arbitrary dependence. The construction also encompasses a fixed-sequence FDR procedure as a special case when the entire initial allocation is assigned to the first hypothesis.

We evaluate the procedures through Gaussian simulations varying signal placement, effect size, dependence structure, and allocation weights. Comparisons with LOND, LORD++ \citep{ramdas2017online}, the procedures of \citet{lynch2017control}, conventional fallback, and the offline BH and BY procedures reveal modest power gains from position-dependent calibration and improvements of PRDS fallback over conventional fallback in the principal independent scenarios. The simulations also illustrate the power loss associated with the arbitrary-dependence correction. A separate experiment investigates the PLTDN-based procedure for two-sided Gaussian testing with known covariance, where calibration incorporates the dependence structure into the testing rule. These comparisons highlight specific advantages and limitations of the proposed methods without suggesting uniform power superiority.

The remainder of the paper is organized as follows.
Section~\ref{s:setup} presents the testing framework, notation, assumptions, and connections with existing methods. Section~\ref{s:procedure} develops generalized fixed-sequence procedures under independence and PLTDN. Section~\ref{sec:fb} introduces the fallback procedures and establishes FDR control at deterministic times within the planned horizon.
Section~\ref{s:sim} presents the simulation results, and
Section~\ref{s:discussion} discusses limitations and directions for future research. Proofs are provided in the appendix.

\section{Problem Formulation and Related Work}
\label{s:setup}

We introduce the ordered testing framework, notation,
and standing assumptions. We then define the two
threshold structures studied in Sections~\ref{s:procedure}
and~\ref{sec:fb} and explain their connections with
existing multiple testing methods.

\subsection{Ordered testing framework}

Consider a fixed, known family of hypotheses
$H_1,\ldots,H_m$, arranged in a prespecified order.
The order may reflect scientific priorities, external
information, or the sequence in which data become
available; it is not obtained by sorting the testing
$p$-values.

Let $P_1,\ldots,P_m$ denote the corresponding
$p$-values. Hypotheses are evaluated sequentially:
the decision for $H_i$ may use $P_1,\ldots,P_i$,
but not future $p$-values. Each decision is
irrevocable, meaning that its rejection status
is not subsequently revised.

The principal procedures allow testing to continue
after non-rejections. The theoretical results concern
a fixed horizon $m$; unknown horizons and
data-dependent termination are not automatically
covered by these guarantees.

\subsection{Notation and error criteria}

Let
\[
I_0=\{i:H_i\text{ is true}\},
\qquad
I_1=\{1,\ldots,m\}\setminus I_0,
\]
and write $m_0=|I_0|$ and $m_1=m-m_0$.
These sets and their sizes are unknown.

Define the preceding testing history by
\[
\boldsymbol P_{<i}=(P_1,\ldots,P_{i-1}),
\qquad
\mathcal F_{i-1}=\sigma(P_1,\ldots,P_{i-1}),
\]
with $\mathcal F_0$ the trivial sigma-field.
A sequential threshold $T_i\in[0,1]$ is
\emph{predictable} if it is
$\mathcal F_{i-1}$-measurable: its value is determined
before observing $P_i$.

The rejection indicator is
\[
D_i=\mathbf1\{T_i>0,\ P_i\le T_i\}.
\]
Thus, a zero threshold produces no rejection,
including when $P_i=0$.
This convention is relevant when zero allocations
are used to implement a stopping rule.

Let
\[
R_i=\sum_{j=1}^{i}D_j,
\qquad
V_i=\sum_{\substack{j\le i\\j\in I_0}}D_j,
\qquad
S_i=\sum_{\substack{j\le i\\j\in I_1}}D_j,
\]
with $R_0=V_0=S_0=0$.
These count total rejections, false discoveries,
and true discoveries among the first $i$ hypotheses,
respectively, and satisfy $R_i=V_i+S_i$.

The false discovery proportion and false discovery
rate at position $t$ are
\[
\mathrm{FDP}(t)=\frac{V_t}{R_t\vee1},
\qquad
\mathrm{FDR}(t)=
\mathbb E\left[\frac{V_t}{R_t\vee1}\right],
\]
where $a\vee b=\max\{a,b\}$.
Unless stated otherwise, we write
$\mathrm{FDR}=\mathrm{FDR}(m)$. The main theorems establish
terminal control at a prespecified level $\alpha\in(0,1)$
\citep{Benjamini_1995}; Corollary~\ref{cor:prefix-fdr}
extends these guarantees to every deterministic time
$t\le m$.

For comparison, the terminal family-wise error rate is
\[
\mathrm{FWER}=\Pr(V_m\ge1).
\]
Since $\mathrm{FDP}(m)\le\mathbf1\{V_m\ge1\}$,
FWER control implies FDR control at the same level.

All FDR guarantees are required to hold under every
configuration of true and false null hypotheses
and every joint distribution satisfying the
assumptions of the corresponding result.

\subsection{Threshold structures and stopping rules}
\label{subsec:procedural-structures}

We distinguish two threshold structures according
to their use of preceding rejection decisions.

\begin{definition}[Rejection-count generalized fixed-sequence procedure]
\label{def:gfs-general}
A rejection-count generalized fixed-sequence procedure
uses thresholds
\[
T_i=\widetilde\alpha_i(R_{i-1}),
\qquad i=1,\ldots,m,
\]
where each
$\widetilde\alpha_i:\{0,\ldots,i-1\}\to[0,1]$
is deterministic.
\end{definition}

At a given testing position, histories with the
same rejection count yield the same threshold.
The position $i$ also determines the number of
preceding non-rejections, $i-1-R_{i-1}$.
This structure is related to the generalized
fixed-sequence framework of
\citet{qiu2015generalized}.
Section~\ref{s:procedure} studies a
position-dependent rule under independence and
a PLTDN-calibrated linear rule.

Fallback-type procedures retain additional information
through transferred testing levels.
Let deterministic weights $\omega_i\ge0$ satisfy
$\sum_{i=1}^{m}\omega_i=1$, and define
\begin{equation}
L_1=\alpha\omega_1,
\qquad
L_i=\alpha\omega_i+D_{i-1}L_{i-1},
\quad i=2,\ldots,m.
\label{eq:setup-fb-transfer}
\end{equation}

\begin{definition}[Fallback-type procedure]
\label{def:fb-general}
A fallback-type procedure uses the levels
in \eqref{eq:setup-fb-transfer} and thresholds
\[
T_i=\min\{1,h_i(R_{i-1})L_i\},
\]
where
$h_i:\{0,\ldots,i-1\}\to[0,\infty)$
is deterministic.
\end{definition}

The recursion carries allocations forward after
rejections and resets to $\alpha\omega_i$
after a non-rejection. Thus, histories with the
same rejection count can yield different thresholds.
Section~\ref{sec:fb} studies a particular multiplier
and its arbitrary-dependence correction.

These definitions describe the constructions examined
here, rather than an exhaustive or necessarily disjoint
classification of sequential procedures.
Both structures are predictable.
The transfer recursion is established in conventional
fallback testing and its online extensions
\citep{Wiens_2003,Wiens_2005,tian2021online}.

Stopping behavior is a separate component.
We distinguish:
\begin{itemize}
\item \emph{Fixed-horizon continuation:}
the procedure evaluates all $m$ hypotheses and
may continue after any non-rejection.

\item \emph{Non-rejection-based stopping:}
testing terminates when a prespecified number
$K$ of non-rejections has occurred.
The termination position can be written as
\[
\tau_K
=
\min\left(
\{i\in\{1,\ldots,m\}:i-R_i\ge K\}
\cup\{m\}
\right),
\]
with thresholds set to zero after termination.
Conventional fixed-sequence testing corresponds
to $K=1$.
\end{itemize}

The principal procedures use fixed-horizon continuation.
Certain zero-weight choices in the fallback construction
recover the rejection set of a stopping
fixed-sequence procedure.
Such a special case does not imply validity under
general data-dependent stopping.

\subsection{Marginal validity and dependence conditions}
\label{subsec:dependence}

Unless explicitly replaced by an alternative
marginal calibration, true-null $p$-values satisfy
\begin{equation}
\Pr(P_i\le u)\le u,
\qquad u\in[0,1],\quad i\in I_0.
\label{eq:marginal-validity}
\end{equation}
This is a marginal assumption; conditional
super-uniformity given $\mathcal F_{i-1}$
is not a standing requirement.

For the PLTDN result, $P_i$ may instead denote a
transformed testing quantity with a nonuniform
null distribution. In that case, the result
specifies deterministic bounds
\[
\Pr(P_i\le u)\le F_i(u),
\qquad i\in I_0,
\]
in place of \eqref{eq:marginal-validity}.
The bounds must be justified without knowledge
of $I_0$.

We consider mutual independence, arbitrary dependence,
and two history-based positive-dependence conditions.

\begin{definition}[History-based PRDS]
\label{def:history-prds}
The ordered testing quantities satisfy history-based
positive regression dependence on a subset
(PRDS) if, for each $i\in I_0$ and every bounded,
coordinatewise nondecreasing measurable function
\[
h:[0,1]^{i-1}\to\mathbb R,
\]
the function
\[
u\longmapsto
\mathbb E\left[
h(\boldsymbol P_{<i})\mid P_i\le u
\right]
\]
is nondecreasing over values $u\in(0,1]$
for which $\Pr(P_i\le u)>0$.
\end{definition}

This uses a cumulative-conditioning formulation
of PRDS, restricted to preceding testing quantities;
see \citet{Benjamini_2001,Finner_2007,blanchard2008two}.

\begin{definition}[History-based PLTDN]
\label{def:history-pltdn}
The ordered testing quantities satisfy history-based
positive left-tail dependence under the null
(PLTDN) if, for each $i\in I_0$, every coordinatewise
nondecreasing measurable function
\[
h:[0,1]^{i-1}\to\mathbb R,
\]
and every $c\in\mathbb R$ such that
$\Pr\{h(\boldsymbol P_{<i})\le c\}>0$,
the function
\[
t\longmapsto
\frac{
\Pr\{P_i\le t\mid h(\boldsymbol P_{<i})\le c\}
}{t}
\]
is nonincreasing on $(0,1]$.
\end{definition}

This is the history-based version of the PLTDN
condition introduced by \citet{GhoshSarkar2025}.
For $i=1$, history-based PRDS is vacuous,
whereas history-based PLTDN requires
$\Pr(P_1\le t)/t$ to be nonincreasing.

Both definitions concern the relationship between
each true-null quantity and its predecessors.
They impose no analogous requirement involving
future testing quantities.
For exactly uniform null marginals, the two
history-based conditions are equivalent, as
explained in Section~\ref{s:Depend}.
For nonuniform marginals, their respective
calibration requirements must be considered separately.

Under independence, $P_1,\ldots,P_m$ are mutually
independent. Under arbitrary dependence, no joint
distributional restriction is imposed beyond
the stated marginal conditions.
Independence with exactly uniform null marginals
satisfies both positive-dependence conditions;
independence and marginal super-uniformity
alone do not ensure PLTDN.

\subsection{Technical connections with online testing}
\label{subsec:online-connections}

The literature reviewed in Section~\ref{s:intro} shares
the sequential information constraint of the present
framework. The relevant comparisons concern threshold
construction, dependence assumptions, and the scope
of the error-control guarantee.

\paragraph{Rejection-count and history-dependent rules.}
LOND uses thresholds
\[
T_i=\min\{1,(R_{i-1}+1)a_i\},
\qquad \sum_i a_i\le\alpha,
\]
under the assumptions supporting its stated guarantee
\citep{javanmard2015online}.
Application to a finite family does not change this
construction; finite-horizon allocations have also
been studied in platform trials
\citep{zehetmayer2022online}.
Section~\ref{s:procedure} identifies the additional
calibration used by the position-dependent and
PLTDN procedures.

LORD and generalized alpha-investing rules already
use features of the rejection history beyond its
cumulative count \citep{javanmard2018online}.
Likewise, the transfer recursion
\eqref{eq:setup-fb-transfer} is Online Fallback-1
for FWER control \citep{tian2021online}.
The fallback procedures in Section~\ref{sec:fb}
retain this recursion and add a remaining-horizon
multiplier, transferring the underlying level $L_i$
rather than the enlarged testing threshold.

\paragraph{Scope of the guarantees.}
The FDR defined in Section~\ref{s:setup} is the same
error-rate functional used in online FDR testing.
Corollary~\ref{cor:prefix-fdr} shows that the four
principal constructions satisfy
$\mathrm{FDR}(t)\le\alpha$ at every deterministic
time $t\le m$, while retaining their planned horizon
$m$. This is a collection of expected-FDP bounds,
not a bound on the probability that the FDP crosses
a boundary at some time. It also does not establish
FDR control at arbitrary data-dependent stopping
times or provide a rule for an unknown horizon.
Table~\ref{tab:procedure-summary} summarizes the
principal constructions after both classes have
been introduced.

\section{Generalized Fixed-Sequence Procedures}
\label{s:procedure}

We first present a position-dependent rejection-count
procedure under independence, followed by a sequential
PLTDN calibration for linear thresholds. Connections
with established LOND procedures are discussed in remarks.
Throughout, we use the framework, notation, and standing
assumptions of Section~\ref{s:setup}. Proofs are collected
in the appendix.

\subsection{A Position-Dependent Procedure under Independence}
\label{subsec:gfs-indep}

Suppose that $P_1,\ldots,P_m$ are mutually independent.
The following procedure uses the known family size $m$,
the testing position, and the number of preceding
rejections to determine each rejection threshold.

\begin{procedure}[Position-dependent rejection-count procedure]
\label{proc:gfs-indep}
For $i=1,\ldots,m$, reject $H_i$ whenever
\[
P_i\le\widetilde{\alpha}_i(R_{i-1}),
\]
where
\begin{equation}
\widetilde{\alpha}_i(r)
=
\frac{(r+1)\alpha}
     {i\alpha+(m-r)(1-\alpha)},
\qquad r=0,\ldots,i-1.
\label{eq:gfs-position}
\end{equation}
Testing continues through the entire sequence,
regardless of preceding non-rejections.
\end{procedure}

For $\alpha\in(0,1)$, these thresholds lie strictly
between zero and one. They increase with the preceding
rejection count $r$ and, for fixed $r$, decrease with
the testing position $i$. Thus, discoveries increase
subsequent testing levels, whereas advancing through
the sequence without additional discoveries reduces them.

\begin{theorem}[FDR control under independence]
\label{thm:position-indep}
Under mutual independence and the standing
marginal-validity assumption,
Procedure~\ref{proc:gfs-indep} controls the FDR
at level $\alpha$.
\end{theorem}

The proof is given in Appendix~\ref{APP:A6}.
It uses a decomposition indexed by successive rejections,
following \citet{lynch2017control}, together with a
potential-function argument for the thresholds
in \eqref{eq:gfs-position}.

\begin{remark}[Two benchmark rejection histories]
\label{rem:position-benchmarks}
Two extreme histories illustrate how preceding
discoveries affect the rejection thresholds.

If no preceding hypothesis has been rejected, then
\[
\widetilde{\alpha}_i(0)
=
\frac{\alpha}{i\alpha+m(1-\alpha)}.
\]
These thresholds decrease with $i$ and satisfy
\[
\frac{\alpha}{m}
\le
\widetilde{\alpha}_i(0)
\le
\frac{\alpha}{m-(m-1)\alpha}.
\]
Thus, before the first discovery, the testing levels
are slightly larger than the Bonferroni level
$\alpha/m$ when $\alpha$ is small, with equality
at the final position.

If all preceding hypotheses have been rejected,
then $R_{i-1}=i-1$ and
\[
\widetilde{\alpha}_i(i-1)
=
\frac{i\alpha}
     {i\alpha+(m-i+1)(1-\alpha)}
\ge
\frac{i\alpha}{m}.
\]
Along this history, the thresholds increase with $i$
and can substantially exceed the BH-type reference
levels $i\alpha/m$. In particular,
\[
\widetilde{\alpha}_m(m-1)
=
\frac{m\alpha}{m\alpha+1-\alpha}
\longrightarrow 1
\qquad\text{as }m\to\infty
\]
for fixed $\alpha\in(0,1)$.

The procedure therefore begins with testing levels
close to Bonferroni and permits substantially larger
levels when discoveries accumulate.
Here, the BH-type levels serve only as numerical
benchmarks: $i$ denotes the prespecified testing
position, rather than the rank of an observed
$p$-value.
\end{remark}

\begin{remark}[Comparison with existing rejection-count rules]
\label{rem:position-fixed-sequence}
\label{rem:position-thresholds}
The fixed-sequence procedure of
\citet[Theorem~4.2]{lynch2017control}
stops at a prespecified $K$th non-rejection and uses
critical values
\[
\frac{(r+1)\alpha}
     {i\alpha+K(1-\alpha)},
\]
where $r$ is the number of preceding rejections.
Taking $K=m$ permits evaluation of the entire family.
Procedure~\ref{proc:gfs-indep} instead uses $m-r$
in the denominator, increasing the thresholds after
discoveries. The validity of this history-dependent
replacement requires the argument in
Theorem~\ref{thm:position-indep}; it does not follow
directly from the fixed-$K$ result.

For comparison, the original LOND rule
\citep{javanmard2015online}, applied to a fixed family
with equal allocations, uses thresholds
\[
T_i^{\mathrm{LOND}}
=
\min\left\{1,(R_{i-1}+1)\frac{\alpha}{m}\right\}.
\]
For every admissible $(i,r)$,
\[
\widetilde{\alpha}_i(r)
\ge\frac{(r+1)\alpha}{m}.
\]
Because the thresholds are nondecreasing in the
preceding rejection count, induction shows that
Procedure~\ref{proc:gfs-indep} rejects every hypothesis
rejected by equal-allocation LOND on the same
$p$-value sequence. The same argument establishes
containment of the rejection set of the fixed-$K$
procedure with $K=m$. These comparisons do not imply
dominance over other LOND allocations or other
fixed-$K$ choices.
\end{remark}

\begin{remark}[Connection with adaptive step-down testing]
\label{rem:gbs-connection}
The thresholds in \eqref{eq:gfs-position} are also
related to the adaptive step-down critical values of
\citet{gavrilov2009adaptive},
\[
c_k
=
\frac{k\alpha}{m+1-k(1-\alpha)}
=
\frac{k\alpha}{k\alpha+(m-k+1)}.
\]
Writing $k=R_{i-1}+1$, our threshold can be expressed as
\[
\widetilde{\alpha}_i(k-1)
=
\frac{k\alpha}
{k\alpha+(m-k+1)-\alpha(m-i+1)}.
\]
It therefore exceeds $c_k$ at the same index $k$.

The procedures nevertheless use different information
and decision structures. Gavrilov et al.\ apply their
constants to the sorted $p$-values and stop at the first
failed comparison. Our procedure tests hypotheses in
a prespecified order, uses only currently available
$p$-values, and continues after non-rejections.
Consequently, the comparison of critical values does
not imply an ordering of power, and their step-down
FDR theorem does not directly establish
Theorem~\ref{thm:position-indep}.
\end{remark}

\begin{remark}[Exact global-null calibration]
\label{rem:position-global-null}
Under the global null with independent, exactly uniform
$p$-values,
\begin{align*}
\Pr(R_m=0)
&=
\prod_{i=1}^{m}
\left\{
1-\frac{\alpha}{i\alpha+m(1-\alpha)}
\right\}\\
&=
\prod_{i=1}^{m}
\frac{m(1-\alpha)+(i-1)\alpha}
     {m(1-\alpha)+i\alpha}
=1-\alpha.
\end{align*}
Thus, under the global null, the FDR equals
$\Pr(R_m>0)=\alpha$, showing that the bound in
Theorem~\ref{thm:position-indep} is attained.
\end{remark}

\subsection{Sequential Calibration under PLTDN}
\label{s:Depend}

We now extend linear rejection-count testing to
the history-based PLTDN setting of
Definition~\ref{def:history-pltdn}.
The extension combines this dependence condition
with calibration through null marginal distribution
functions, allowing the tested quantities to have
nonuniform null distributions.

For each $i=1,\ldots,m$, let
$F_i:[0,1]\to[0,1]$ be a deterministic,
nondecreasing function satisfying
\[
\Pr(P_i\le u)\le F_i(u),
\qquad u\in[0,1],\quad i\in I_0.
\]
These bounds replace the standing marginal-validity
assumption in this subsection's PLTDN result.
The functions $F_i$ may be exact null distribution
functions or justified deterministic upper bounds,
specified without knowledge of $I_0$.

\begin{procedure}[PLTDN-calibrated linear procedure]
\label{proc:gfs-pltdn}
Choose deterministic allocations
$a_1,\ldots,a_m\in[0,1]$ satisfying
\[
\sum_{i=1}^{m}F_i(a_i)\le\alpha.
\]
For $i=1,\ldots,m$, set
\begin{equation}
T_i=\min\{1,(R_{i-1}+1)a_i\}.
\label{eq:gfs-linear}
\end{equation}
Reject $H_i$ if $T_i>0$ and $P_i\le T_i$.
Testing continues through the entire sequence,
regardless of preceding non-rejections.
\end{procedure}

Procedure~\ref{proc:gfs-pltdn} retains the linear
rejection-count structure of LOND
\citep{javanmard2015online}, but changes both the
dependence assumption and the allocation requirement.
Compared with the PRDS-based guarantee under
super-uniform null marginals and
$\sum_i a_i\le\alpha$, the following result uses
history-based PLTDN and the calibrated budget
$\sum_iF_i(a_i)\le\alpha$.

\begin{theorem}[FDR control under history-based PLTDN]
\label{thm:FDR_PLTDN}
Suppose that the ordered testing quantities satisfy
Definition~\ref{def:history-pltdn} and the marginal
bounds specified above. Then
Procedure~\ref{proc:gfs-pltdn} controls the FDR
at level $\alpha$.
\end{theorem}

The proof is given in Appendix~\ref{app:proof-pltdn}.
It extends the PLTDN argument of
\citet{GhoshSarkar2025} to the sequential setting:
the contribution associated with each true null
is bounded by
\[
\Pr(P_i\le a_i)\le F_i(a_i),
\]
and summing these bounds establishes FDR control.
The PLTDN concept and the underlying
marginal-calibration principle originate in that
work; the present result applies them to sequential
rejection-count testing with hypothesis-specific
allocations.

\begin{remark}[Choice of allocations]
\label{rem:positive-allocation}
For a common allocation $a_i=a$, the budget condition
becomes
\[
\overline{F}(a)\le\frac{\alpha}{m},
\qquad
\overline{F}(u)=\frac1m\sum_{i=1}^{m}F_i(u).
\]
If $\overline{F}$ is continuous and strictly increasing
and $\alpha/m$ belongs to its range, one may choose
\[
a=\overline{F}^{-1}(\alpha/m).
\]
Otherwise, the selected allocation must be checked
directly against the budget inequality.
Unequal allocations may reflect prespecified
scientific priorities, provided that
$\sum_iF_i(a_i)\le\alpha$.
\end{remark}

\begin{remark}[Applications to transformed testing quantities]
\label{rem:history-dependence}
The marginal calibration is relevant to certain
dependence-adjusted transformations in two-sided
Gaussian testing
\citep{Sarkar_2025,GhoshSarkar2025}.
Such transformations can yield quantities satisfying
PLTDN whose null distributions are not uniform.
Their distribution functions then determine the
allocation budget in
Procedure~\ref{proc:gfs-pltdn}.

Application of the theorem requires verifying PLTDN
and the marginal bounds for the quantities actually
used in testing. Lemma~\ref{lem:gaussian-history-pltdn} in
Appendix~\ref{app:gaussian-pltdn} verifies these
requirements for the Gaussian construction used
in Section~\ref{s:sim}. Data-dependent transformations
or estimated calibration functions require
additional validity arguments.
\end{remark}

\begin{remark}[Relation to history-based PRDS]
\label{rem:dependence-interpretation}
PLTDN and PRDS impose different conditional
monotonicity requirements in general.
For exactly uniform null marginals, however,
Bayes' rule gives
\[
\frac{\Pr(P_i\le t\mid A)}{t}
=
\frac{\Pr(A\mid P_i\le t)}{\Pr(A)},
\qquad t\in(0,1],
\]
for every decreasing history event $A$ with
$\Pr(A)>0$.
Consequently, the history-based PLTDN and PRDS
conditions defined in
Section~\ref{subsec:dependence} are equivalent
in that case.

The PLTDN extension is therefore particularly
relevant for nonuniform null quantities, where
the dependence condition and marginal calibration
must be considered together. It is not asserted
that PLTDN is universally weaker than PRDS.
\end{remark}

\begin{remark}[Established PRDS calibration of LOND]
\label{thm:FDR_PRDS}
\label{rem:lond-bh}
Under the standing marginal-validity assumption
and history-based PRDS, the linear rule
\eqref{eq:gfs-linear} controls the FDR whenever
\[
\sum_{i=1}^{m}a_i\le\alpha.
\]
This follows from the dependency-control inequality
of \citet[Lemma~3.2(ii)]{blanchard2008two}, applied
to $(P_i,R_{i-1}+1)$; see
Appendix~\ref{app:proof-prds}.
For PRDS validity of LOND in the online-testing
literature, see also \citet{xu2024online}.
\end{remark}

\begin{remark}[Established arbitrary-dependence calibration]
\label{s:ArbitDep}
\label{thm:FDRArbDep}
Under marginal validity and arbitrary dependence,
the dependence-adjusted LOND rule
\[
T_i
=
\min\left\{
1,\frac{(R_{i-1}+1)a_i}{H_i}
\right\},
\qquad
H_i=\sum_{k=1}^{i}\frac1k,
\]
controls the FDR whenever $\sum_i a_i\le\alpha$
\citep[Theorem~2.7]{javanmard2015online}.
A verification in the present notation is given
in Appendix~\ref{app:proof-harmonic}.
More general calibrations follow from established
reshaping inequalities \citep{blanchard2008two}.
These serve as benchmarks for the procedures
developed here.
\end{remark}

\section{Fallback-Type Procedures for Controlling the FDR}
\label{sec:fb}

We now consider procedures that combine rejection-count
adaptation with recursive level transfer.
Unlike the generalized fixed-sequence rules of
Section~\ref{s:procedure}, these thresholds depend on
the arrangement of preceding rejections as well as
their cumulative number.

Let $\omega_1,\ldots,\omega_m$ be prespecified,
nonnegative weights satisfying
$\sum_{i=1}^{m}\omega_i=1$.
Define the transferred level by
\begin{equation}
L_1=\alpha\omega_1,
\qquad
L_i=\alpha\omega_i+D_{i-1}L_{i-1},
\quad i=2,\ldots,m.
\label{eq:fb-transfer}
\end{equation}
Equivalently,
\[
L_i
=
\alpha\sum_{j=1}^{i}\omega_j
\prod_{\ell=j}^{i-1}D_\ell,
\]
where an empty product equals one.
An allocation is therefore transferred forward
along consecutive rejections and is interrupted
by a non-rejection.

The recursion in \eqref{eq:fb-transfer} is the
conventional fallback mechanism
\citep{Wiens_2003,Wiens_2005}.
It also appears as Online Fallback-1 in
\citet{tian2021online}, where it provides
FWER control under arbitrary dependence.
Our construction retains this mechanism and introduces
an additional multiplier for FDR control.
We study the resulting rule under history-based PRDS
and a corrected version under arbitrary dependence.

\subsection{Fallback-Type Procedure under PRDS}
\label{subsec:fb-prds}

For $i=1,\ldots,m$, write
\[
h_i(r)=1+\frac{r}{m-i+1},
\qquad r=0,\ldots,i-1.
\]
The multiplier increases with the number of preceding
discoveries and uses the number of remaining hypotheses,
including the current one.

\begin{procedure}[Fallback-type procedure under PRDS]
\label{proc:fb-prds}
At position $i$, compute $L_i$ from
\eqref{eq:fb-transfer} using this procedure's
preceding decisions, and set
\[
T_i
=
\min\left\{
1,\left(1+\frac{R_{i-1}}{m-i+1}\right)L_i
\right\}.
\]
Reject $H_i$ if $T_i>0$ and $P_i\le T_i$.
Otherwise, do not reject $H_i$ and proceed to
the next hypothesis.
\end{procedure}

\begin{theorem}[FDR control under history-based PRDS]
\label{thm:fb-prds}
Under the standing marginal-validity assumption
and history-based PRDS in
Definition~\ref{def:history-prds},
Procedure~\ref{proc:fb-prds} controls the FDR
at level $\alpha$.
\end{theorem}

The proof is given in Appendix~\ref{APP:A7}.
It decomposes the FDR according to the location
of the first rejected true null, following the
type of argument used by \citet{lynch2017control}.
On each such event, the transferred level is bounded
by a deterministic sum of allocations.
A positive-dependence inequality then bounds the
corresponding FDR contribution.

\begin{remark}[Comparison with conventional fallback]
\label{rem:fb-dependence}
Conventional fallback uses the same recursion
but tests at level $L_i$.
Since $h_i(r)\ge1$, monotonicity of the thresholds
and transfer recursion shows that
Procedure~\ref{proc:fb-prds} contains the conventional
fallback rejection set for every $p$-value sequence,
with the same weights.

The guarantees differ: conventional fallback
controls the FWER under arbitrary dependence,
whereas Theorem~\ref{thm:fb-prds} establishes FDR
control under history-based PRDS.
The FWER validity of the transfer recursion alone
does not justify the additional multiplier.
\end{remark}

\begin{remark}[Fixed-sequence special case]
\label{rem:fb-fixed-sequence}
Suppose $\omega_1=1$ and $\omega_i=0$ for $i\ge2$.
Along the initial uninterrupted sequence of rejections,
$L_i=\alpha$ and $R_{i-1}=i-1$, giving
\[
T_i
=
\min\left\{1,\frac{m\alpha}{m-i+1}\right\}.
\]
After the first non-rejection, all subsequent
transferred levels are zero.
The resulting rejection set therefore coincides
with the fixed-sequence FDR procedure of
\citet{lynch2017control}.

That special case controls the FDR under arbitrary
dependence. Thus, the PRDS condition in
Theorem~\ref{thm:fb-prds} is sufficient for the
general weighted construction; it is not necessary
for every choice of weights.
\end{remark}

\subsection{Fallback-Type Procedure under Arbitrary Dependence}
\label{subsec:fb-arb}

To accommodate arbitrary dependence, retain
\eqref{eq:fb-transfer} and introduce
\[
c_i
=
1+\sum_{k=m-i+2}^{m}\frac1k
=
1+H_m-H_{m-i+1},
\]
where $H_r=\sum_{k=1}^{r}1/k$ and an empty sum is zero.

\begin{procedure}[Corrected fallback-type procedure]
\label{proc:fb-arb}
At position $i$, compute $L_i$ from
\eqref{eq:fb-transfer} using this procedure's
preceding decisions, and set
\[
T_i
=
\min\left\{
1,\left(1+\frac{R_{i-1}}{m-i+1}\right)
\frac{L_i}{c_i}
\right\}.
\]
Reject $H_i$ if $T_i>0$ and $P_i\le T_i$.
Otherwise, do not reject $H_i$ and proceed to
the next hypothesis.
\end{procedure}

\begin{theorem}[FDR control under arbitrary dependence]
\label{thm:fb-arb}
Under the standing marginal-validity assumption,
Procedure~\ref{proc:fb-arb} controls the FDR
at level $\alpha$ under arbitrary dependence.
\end{theorem}

The proof is given in Appendix~\ref{APP:A8}.
It uses the same first-false-rejection decomposition
as the PRDS argument, replacing the positive-dependence
inequality with an arbitrary-dependence bound.
This probability bound is an instance of established
dependency-control and reshaping arguments
\citep{blanchard2008two}.
The factor $c_i$ reflects the range of rejection counts
arising in the fallback decomposition.

\begin{remark}[Interpretation of the correction]
\label{rem:fb-correction}
The factors satisfy
\[
1=c_1\le c_2\le\cdots\le c_m=H_m.
\]
The correction is therefore smaller than a global
$H_m$ correction at every position before the last.
Monotonicity of the thresholds and transfer recursion
implies that Procedure~\ref{proc:fb-arb} contains
the rejection set of the corresponding rule using
$H_m$ at every position.

This comparison concerns two corrections of the
same fallback construction. It does not imply
dominance over other arbitrary-dependence
FDR procedures.
\end{remark}

\begin{remark}[Comparison with conventional fallback]
\label{rem:fb-arbitrary-comparison}
At the same testing history and with $L_i>0$,
the ratio of the untruncated corrected threshold
to the conventional fallback threshold is
\[
\frac{1+R_{i-1}/(m-i+1)}{c_i}.
\]
It is at least one precisely when
\[
R_{i-1}\ge(m-i+1)(c_i-1).
\]
Thus, the corrected rule can use smaller thresholds
when few discoveries have occurred and larger
thresholds when sufficiently many have accumulated.
There is no uniform rejection-set ordering relative
to conventional fallback. Taking the larger of
the two thresholds at each step would require
a separate FDR-control argument.
\end{remark}

\subsection{FDR Control Within the Planned Horizon}
\label{subsec:fb-online}
\label{subsec:prefix-fdr}

The terminal guarantees also yield control at each
deterministic time within the planned sequence.
This extension applies to all four principal
constructions, with the original horizon $m$
retained in their threshold formulas.

\begin{corollary}[FDR control at deterministic testing times]
\label{cor:prefix-fdr}
Fix the planned horizon $m$. Under the respective
assumptions of Theorems~\ref{thm:position-indep},
\ref{thm:FDR_PLTDN}, \ref{thm:fb-prds}, and
\ref{thm:fb-arb}, the corresponding procedure satisfies
\[
\mathrm{FDR}(t)
=\mathbb E\!\left[\frac{V_t}{R_t\vee1}\right]
\le\alpha,
\qquad t=1,\ldots,m.
\]
The thresholds and allocations retain their original
specifications for horizon $m$.
\end{corollary}

The proof is given in Appendix~\ref{app:prefix-fdr}.
For PLTDN, the proof bounds the contributions from
true nulls among the first $t$ hypotheses. For the
other constructions, a padding argument reduces
prefix control to the terminal guarantee.
The corollary does not establish validity at arbitrary
data-dependent stopping times or for an unknown horizon.

\subsection{Illustration and Summary}
\label{subsec:fb-history}

The following example illustrates the distinction
between rejection-count and fallback thresholds.

\begin{example}[Equal rejection counts, different thresholds]
\label{ex:count-versus-history}
Let $m=5$ and $\omega_i=1/5$.
Consider two histories after three tests:
\[
\begin{array}{c|ccc|c}
\text{History}&D_1&D_2&D_3&R_3\\
\hline
A&1&0&1&2\\
B&0&1&1&2
\end{array}
\]
A rejection-count rule assigns the same threshold
to $H_4$ under both histories.
For example, equal-allocation LOND gives $3\alpha/5$.

The transferred levels are instead
\[
L_4^A=\frac{2\alpha}{5},
\qquad
L_4^B=\frac{3\alpha}{5}.
\]
Both histories have multiplier
$1+R_3/(m-4+1)=2$ and correction
$c_4=107/60$.
The fallback thresholds before truncation are
\[
\begin{array}{l|cc}
\text{Rule}&\text{History }A&\text{History }B\\
\hline
\text{PRDS fallback}
&4\alpha/5&6\alpha/5\\[3pt]
\text{Corrected fallback}
&4\alpha/(5c_4)&6\alpha/(5c_4)
\end{array}
\]
The difference arises because history $B$ ends
with a longer sequence of consecutive rejections.

These are comparisons at specified histories;
different procedures need not generate the same
history on a common $p$-value sequence.
\end{example}

Table~\ref{tab:procedure-summary} summarizes the
four principal constructions.
The established LOND and dependence-adjusted LOND
benchmarks are discussed in Section~\ref{s:procedure}.

\begin{table}[htbp]
\centering
\caption{Principal constructions and sufficient conditions
for FDR control.}
\label{tab:procedure-summary}

\begingroup
\small
\setlength{\tabcolsep}{4pt}
\renewcommand{\arraystretch}{1.5}

\begin{tabular}{
>{\raggedright\arraybackslash}p{0.23\linewidth}
>{\raggedright\arraybackslash}p{0.39\linewidth}
>{\raggedright\arraybackslash}p{0.29\linewidth}
}
\hline
\textbf{Construction}
&
\textbf{Threshold}
&
\textbf{Conditions}
\\
\hline

Position-dependent rejection-count
&
$\displaystyle
\frac{k\alpha}
{i\alpha+(m-k+1)(1-\alpha)}$
&
Mutual independence.
Theorem~\ref{thm:position-indep}.
\\[3pt]

PLTDN-calibrated linear rule
&
$\displaystyle ka_i$
&
History-based PLTDN;
$\sum_i F_i(a_i)\le\alpha$.
Theorem~\ref{thm:FDR_PLTDN}.
\\[3pt]

PRDS fallback
&
$\displaystyle
\left(1+\frac{k-1}{m-i+1}\right)L_i$
&
History-based PRDS.
Theorem~\ref{thm:fb-prds}.
\\[3pt]

Corrected fallback
&
$\displaystyle
\left(1+\frac{k-1}{m-i+1}\right)
\frac{L_i}{c_i}$
&
Arbitrary dependence.
Theorem~\ref{thm:fb-arb}.
\\[3pt]

\hline
\end{tabular}

\par\smallskip
\begin{minipage}{0.98\linewidth}
\raggedright
\footnotesize
\textit{Notes.}
Here $k=R_{i-1}+1$,
$c_i=1+H_m-H_{m-i+1}$, and $L_i$ follows
\eqref{eq:fb-transfer} using each procedure's own
decisions. Thresholds are capped at one when needed,
and zero thresholds produce no rejections.
The linear threshold form is LOND; its PLTDN
calibration is the application considered here.
\end{minipage}

\endgroup
\end{table}

\section{Simulation Study}
\label{s:sim}

The simulation study examines how rejection-count adaptation,
recursive level transfer, and dependence calibration affect FDR
and power. We organize the experiments around three questions:
how performance changes with the placement of signals in the
sequence; how dependence corrections affect power; and how
prespecified allocations influence the benefits of ordering.
A separate two-sided experiment examines the PLTDN calibration.
All methods are evaluated on the same simulated data within
each replicate.

\subsection{Methods and implementation}
\label{subsec:sim-methods}

The four principal procedures are the position-dependent rule
under independence (Procedure~\ref{proc:gfs-indep}), the
PLTDN-calibrated linear rule (Procedure~\ref{proc:gfs-pltdn}),
and the PRDS and arbitrary-dependence fallback rules
(Procedures~\ref{proc:fb-prds} and~\ref{proc:fb-arb}).
We examine the PLTDN procedure separately in the two-sided experiment.

The sequential comparators include LOND and its position-specific
harmonic correction, conventional fallback
\citep{Wiens_2003,Wiens_2005}, LORD++
\citep{ramdas2017online}, and two procedures of
\citet{lynch2017control}. The first is the stopping
fixed-sequence rule ($K=1$), which tests at
$\min\{1,m\alpha/(m-i+1)\}$ until the first non-rejection.
The second is the independence-based specialization $K=m$,
which evaluates the whole sequence and uses
\[
T_i^{\mathrm{Lynch}}=
\frac{(R_{i-1}+1)\alpha}{i\alpha+m(1-\alpha)}.
\]
Including both distinguishes gains from continued testing
from gains due to the position-dependent calibration.
With equal allocations, the position-dependent rule contains
the rejection set of Lynch's $K=m$ rule, which in turn contains
that of LOND. We check these inclusions on shared samples.
Conventional fallback is also the online fallback FWER benchmark
considered by \citet{tian2021online}; we do not include it twice.
Bonferroni is an additional FWER benchmark. BH and BY are
offline reference procedures with access to all $p$-values.

For the allocation-based rules, the main experiments use
$a_i=\alpha/m$ and $\omega_i=1/m$. The position-dependent
and stopping fixed-sequence rules use their stated formulas.
For LORD++, we set $w_0=\alpha/2$ and prespecify
\[
\gamma_j=\frac{(1-d)d^{j-1}}{1-d^m},\quad j=1,\ldots,m,
\qquad d=0.99,
\]
with $\gamma_j=0$ outside this range. If $\tau_r$ is the
position of the $r$th rejection, its threshold is
\[
T_i=w_0\gamma_i+(\alpha-w_0)\gamma_{i-\tau_1}
+\alpha\sum_{r\ge2:\,\tau_r<i}\gamma_{i-\tau_r},
\]
where the first-rejection term is absent before any rejection.
This schedule is fixed before simulation; comparisons with
LORD++ are therefore conditional on this allocation choice. All thresholds are capped at
one, and zero thresholds produce no rejection.

\subsection{Simulation design}
\label{subsec:sim-design}

\paragraph{Gaussian model and fixed configurations.}
For each scenario, fix the true-null index set and generate
\[
\boldsymbol X\sim\mathcal N_m(\boldsymbol\mu,\Sigma),
\qquad \Sigma_{ij}=\rho^{|i-j|}.
\]
For the one-sided experiments, $\mu_i=\delta$ at the
$m_1$ nonnull positions and $\mu_i=0$ otherwise. We test
$H_i:\mu_i\le0$ using $P_i=1-\Phi(X_i)$.
The covariance is known and the diagonal variances are one.
The order and configuration are fixed before generating the
Monte Carlo samples; no ordering uses observed $p$-values.

The core grid uses $m=200$, $m_1\in\{10,40\}$,
$\delta\in\{2,3\}$, and $\rho\in\{-0.5,0,0.5\}$.
For every combination, we consider five signal arrangements:
\begin{itemize}
\item \emph{Early:} the first $m_1$ positions are nonnull.
\item \emph{Late:} the last $m_1$ positions are nonnull.
\item \emph{Dispersed:} nonnull positions are approximately
      equally spaced across the sequence.
\item \emph{Clustered:} three groups of approximately equal
      size are centered near positions $0.2m$, $0.5m$, and $0.8m$.
\item \emph{Random fixed:} a subset of size $m_1$ is drawn
      once using a recorded seed and held fixed across replicates.
\end{itemize}
Crossing these factors yields $2\times2\times3\times5=60$
core simulation scenarios. Changing the arrangement while
holding $m_1$ fixed separates signal prevalence from signal
placement and the availability of consecutive rejections.

\paragraph{Additional checks.}
Global-null scenarios use $m\in\{200,500\}$ and
$\rho\in\{-0.5,0,0.5\}$. Four larger-horizon scenarios
use $m=500$, $m_1=50$, $\delta=3$, random fixed or
clustered signals, and $\rho\in\{0,0.5\}$.
A weight-sensitivity block uses $m=200$, $m_1=20$,
$\delta=3$, early, late, or random fixed signals, and
$\rho\in\{0,0.5\}$. It compares equal weights with
prespecified front-loaded weights
\[
\omega_i=\frac{0.98^{i-1}}{\sum_{j=1}^{m}0.98^{j-1}},
\qquad a_i=\alpha\omega_i.
\]
Methods without allocation weights retain their original
specifications. Four further scenarios examine $\rho=0.8$:
random fixed and clustered alternatives with $m=200$,
$m_1=20$, and $\delta=3$, together
with global-null checks at $m=200$ and $m=500$.

\paragraph{Two-sided PLTDN experiment.}
For two-sided tests of $H_i:\mu_i=0$, use raw $p$-values
$P_i=\overline F_{\chi^2_1}(X_i^2)$, with nonnull means
of magnitude $\delta$ and signs determined by position parity.
We examine $m=200$, $m_1=20$, $\delta\in\{2,3\}$,
early, random fixed, or clustered signals, and
$\rho\in\{-0.5,0.5\}$. Two corresponding global-null
scenarios are also included. Three additional scenarios at
$\rho=0$ use random fixed signals at each effect size and
a global-null configuration. They check agreement between
the PLTDN calibration and ordinary LOND under independence.

The known AR(1) covariance gives conditional variances
$\tau_1=1$ and $\tau_i=1-\rho^2$ for $i\ge2$.
Define shifted quantities and their null marginal CDFs by
\[
\widetilde P_i=\overline F_{\chi^2_1}(X_i^2/\tau_i),
\qquad
F_i(t)=\overline F_{\chi^2_1}
\left(\tau_i\overline F_{\chi^2_1}^{-1}(t)\right).
\]
A common allocation $a$ is obtained numerically from
$\sum_i F_i(a)=\alpha$. The PLTDN rule rejects when
$\widetilde P_i\le\min\{1,(R_{i-1}+1)a\}$.
The implementation uses the equivalent raw-$p$ threshold
$F_i(\min\{1,(R_{i-1}+1)a\})$.
Lemma~\ref{lem:gaussian-history-pltdn} verifies history-based
PLTDN and the stated marginal distributions. The proof uses
the noncentral $\chi^2_1$ conditional distribution of
$X_i^2/\tau_i$ and a monotone survival-function ratio, and
then passes from the preceding Gaussian coordinates to the
preceding transformed quantities by conditional expectation.
Thus the budget is calibrated using $F_i(a)$ rather than $a$.
The construction uses known covariance parameters; replacing
them by estimates requires separate justification.

\paragraph{Replications, metrics, and uncertainty.}
In total, the design contains 103 scenarios, each with
$N=10000$ replicates and $\alpha=0.05$. Scenario-specific
random seeds are deterministic functions of a recorded master
seed and the scenario specification. The signal configuration
is fixed across all replicates of its scenario.
For replicate $b$, record
\[
\mathrm{FDP}^{(b)}=\frac{V_m^{(b)}}{R_m^{(b)}\vee1},
\qquad
\mathrm{Power}^{(b)}=\frac{S_m^{(b)}}{m_1}
\quad(m_1>0).
\]
FDR and power estimates are replicate averages; power is
undefined under the global null. Monte Carlo standard errors
are the sample standard deviations divided by $\sqrt N$.
Figures show pointwise approximate 95\% Monte Carlo intervals,
computed as the estimate plus or minus $1.96$ standard errors
and truncated to $[0,1]$. Paired differences and their standard
errors are also reported, using the common samples.
The output additionally records FWER and mean rejection counts.
The reported metrics concern position $m$; the deterministic-time
guarantees in Corollary~\ref{cor:prefix-fdr} are theoretical
results, not conclusions drawn from these terminal estimates.

\paragraph{Scope of the guarantees.}
Independence-based guarantees apply at $\rho=0$.
For the one-sided model, nonnegative $\rho$ supports the
history-based PRDS guarantees. Negative correlation provides a robustness check for the
one-sided PRDS rules. For the position-dependent, Lynch $K=m$, and LORD++ rules,
we invoke only their independence guarantees; their evaluations
at $\rho\ne0$ are treated as robustness checks. For the correlated two-sided model, raw-$p$ PRDS
and independence guarantees are not invoked; the shifted
PLTDN calibration is assessed separately. Every result row
identifies whether its stated theoretical assumptions hold.
Simulations under assumptions not covered by a theorem are
robustness checks, not evidence of an extended guarantee.

\subsection{Results}
\label{subsec:sim-results}

\begin{figure}[htbp]
\centering
\includegraphics[width=\linewidth]{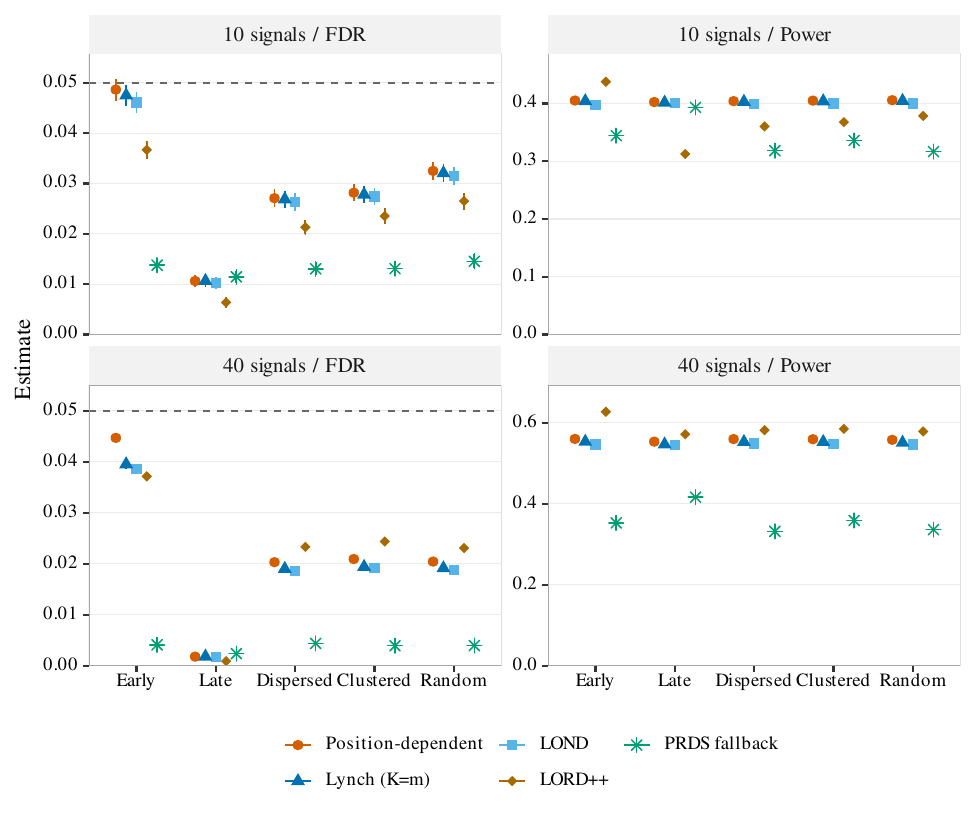}
\caption{Estimated FDR and power under independence for
$m=200$, effect size $3$, and two signal counts. Signal
arrangements are fixed before generating the data. Bars
show pointwise approximate 95\% Monte Carlo intervals.
The dashed line marks the nominal FDR level $0.05$.}
\label{fig:sim-ordering}
\end{figure}

\paragraph{Ordering and independence.}
Figure~\ref{fig:sim-ordering} compares representative sequential
methods under independence at effect size $\delta=3$, including
Lynch's $K=m$ rule. In the
random fixed configuration with $m=200$ and $m_1=40$, the estimated
powers of the linear, position-dependent, and PRDS fallback rules
are $0.547$, $0.558$,
and $0.336$, respectively.
LORD++ achieves $0.579$. The complete results also
include offline BH, with estimated power $0.713$.
Lynch's $K=m$ rule achieves power $0.551$, so the
position-dependent rule gains $0.71$ percentage points
(paired Monte Carlo standard error $0.015$ percentage points).
This direct comparison isolates the gain from the modified
denominator; it is modest in this configuration. Across the
20 independent core scenarios, the position-dependent procedure
has larger estimated power than Lynch's $K=m$ rule in every
case, with gains ranging from less than $0.01$ to $0.71$
percentage points. This agrees with the samplewise rejection-set
inclusion, but does not imply superiority to LORD++ or to
other allocation schedules.

Signal arrangement affects level transfer. At the same signal
count and effect size, estimated PRDS fallback power is
$0.416$ with late signals and
$0.332$ with dispersed signals; classical fallback
gives $0.349$ and
$0.315$, respectively. PRDS fallback has larger estimated
power than conventional fallback in all 20 independent core
scenarios; the largest gain is $6.79$ percentage points.
Both the larger thresholds near the end of the sequence and
accumulation along consecutive rejections contribute to this
sensitivity. These comparisons change the signal arrangement
rather than isolating either mechanism alone. Complete results
include the stopping fixed-sequence comparator, whose low power
in configurations beginning with nulls contrasts with the rules
that allow continued testing.

\begin{figure}[htbp]
\centering
\includegraphics[width=\linewidth]{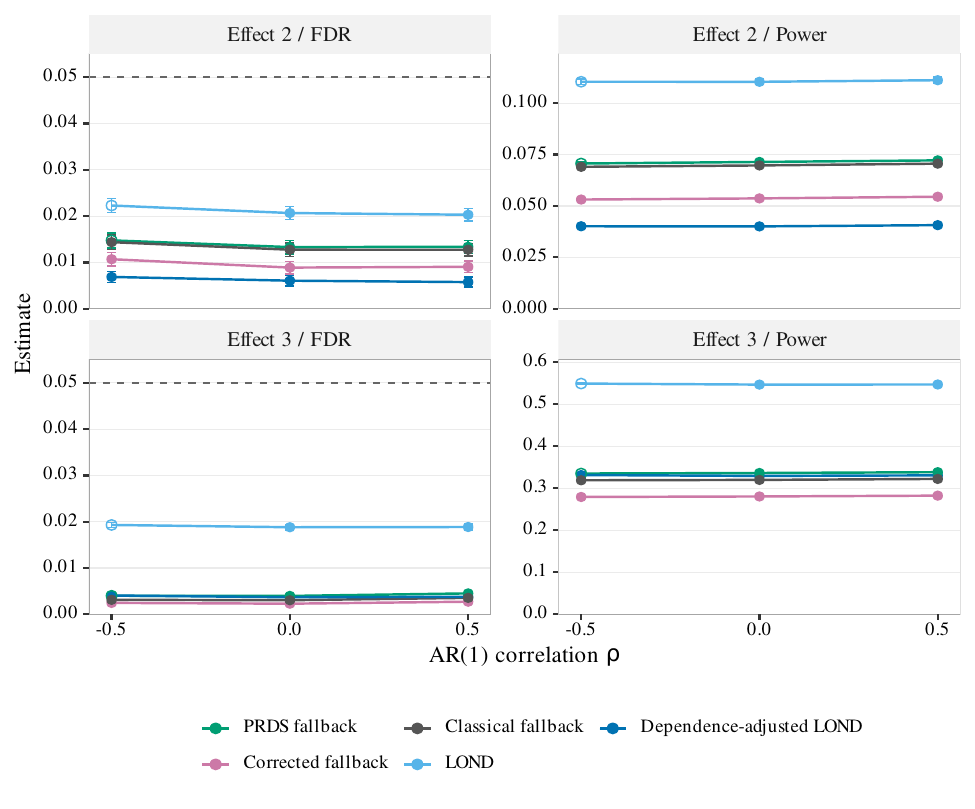}
\caption{Dependence sensitivity with 40 signals among 200
hypotheses in a random fixed arrangement. Filled markers
indicate scenarios covered by the stated guarantee; hollow
markers indicate robustness evaluations outside those
assumptions. The dependence-adjusted LOND and corrected
fallback procedures retain their guarantees at every plotted
correlation. The stated PRDS-based guarantees are not
invoked at negative correlation.}
\label{fig:sim-dependence}
\end{figure}

\paragraph{Dependence calibration.}
Figure~\ref{fig:sim-dependence} shows the trade-off between
power and protection against dependence. The harmonic and
corrected fallback rules are conservative in these scenarios.
The corrected fallback procedure need not improve on classical
fallback: under independence with 40 late signals and effect size
3, its estimated power is $0.265$, compared
with $0.349$ for classical fallback.
This is consistent with the absence of a uniform threshold ordering.

Across the 935 method--scenario combinations covered by
the invoked guarantees, no estimated FDR exceeds $\alpha$
by more than three estimated Monte Carlo standard errors.
This is a descriptive check of numerical compatibility with
the theory, not a simultaneous confidence statement or a
test establishing FDR control.

\paragraph{Global-null calibration.}
Table~\ref{tab:sim-global-null} reports the independent
global-null results, for which FDR equals FWER.
The position-dependent procedure has exact FDR $\alpha$
under independent uniform nulls
(Remark~\ref{rem:position-global-null}). Its estimates
$0.0511$ and $0.0475$ at $m=200$ and $m=500$ are within
approximately $1.2$ Monte Carlo standard errors of $0.05$.
Some estimates coincide because the corresponding procedures
use the same thresholds before the first rejection; under the
global null, FDR depends only on whether any rejection occurs.
The corrected fallback and LORD++ rules are more conservative
in these configurations. These checks
do not establish control over unexamined configurations
or dependence structures.

\begin{table}[htbp]
\centering
\caption{Estimated FDR under the independent global null.
Parentheses contain Monte Carlo standard errors;
$N=10000$ and $\alpha=0.05$. Allocation-based procedures
use equal weights.}
\label{tab:sim-global-null}
\small
\setlength{\tabcolsep}{8pt}
\renewcommand{\arraystretch}{1.2}
\begin{tabular}{lcc}
\hline
\textbf{Procedure} & $m=200$ & $m=500$ \\
\hline
Position-dependent & $0.0511\;(0.0022)$ & $0.0475\;(0.0021)$ \\
Lynch ($K=m$) & $0.0511\;(0.0022)$ & $0.0475\;(0.0021)$ \\
LOND & $0.0493\;(0.0022)$ & $0.0465\;(0.0021)$ \\
PRDS fallback & $0.0493\;(0.0022)$ & $0.0465\;(0.0021)$ \\
Corrected fallback & $0.0304\;(0.0017)$ & $0.0277\;(0.0016)$ \\
Conventional fallback & $0.0493\;(0.0022)$ & $0.0465\;(0.0021)$ \\
LORD++ & $0.0255\;(0.0016)$ & $0.0232\;(0.0015)$ \\
\hline
\end{tabular}
\end{table}

\begin{figure}[htbp]
\centering
\includegraphics[width=\linewidth]{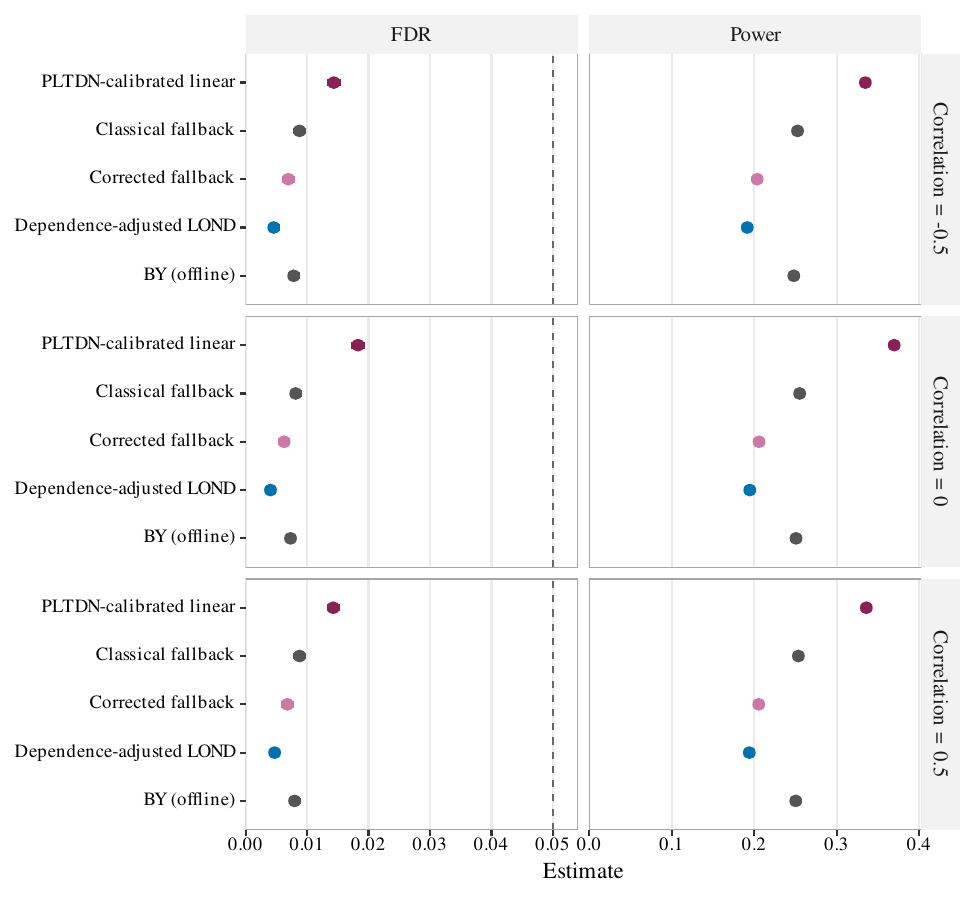}
\caption{Two-sided Gaussian experiment with 20 signals among
200 hypotheses in a random fixed arrangement, with effect size 3. The shifted
PLTDN procedure is compared with methods having
arbitrary-dependence guarantees, at correlations $-0.5$, $0$,
and $0.5$. BY is an offline reference.}
\label{fig:sim-pltdn}
\end{figure}

\paragraph{Two-sided calibration and weights.}
Figure~\ref{fig:sim-pltdn} examines the shifted PLTDN rule.
For random fixed signals of magnitude 3, its estimated power is
$0.335$ at $\rho=-0.5$ and
$0.336$ at $\rho=0.5$, with estimated FDR
$0.014$ and
$0.014$, respectively.
At $\rho=0$, the shifted quantities reduce to the ordinary
two-sided $p$-values, $F_i(t)=t$, and $a=\alpha/m$.
The PLTDN rule then agrees exactly with LOND in all three
independence checks. The figure compares complete procedures, including their different
marginal calibrations, rather than asserting universal power dominance.

\begin{figure}[htbp]
\centering
\includegraphics[width=\linewidth]{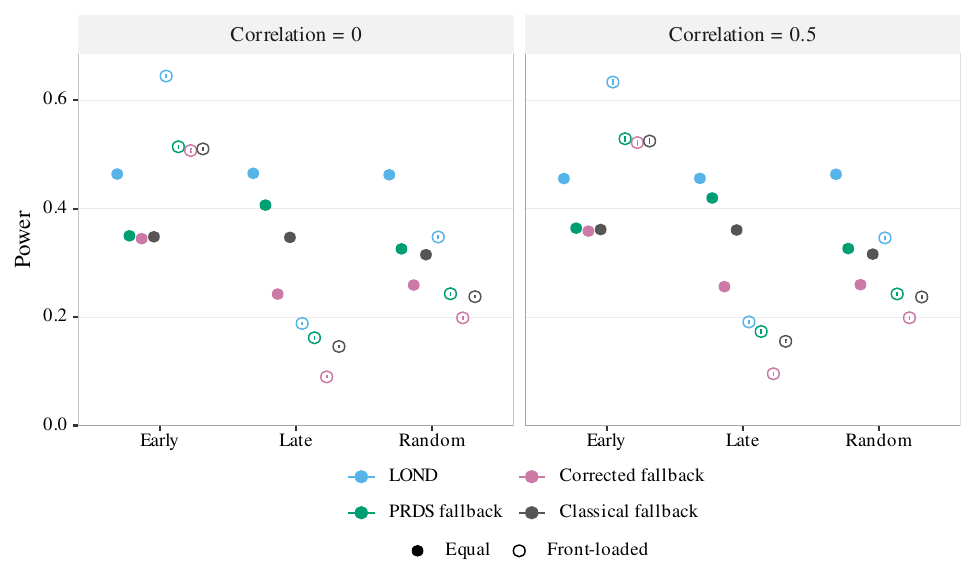}
\caption{Power sensitivity to equal and front-loaded
allocations, with 20 signals among 200 hypotheses and effect
size 3. Filled circles denote equal weights and open circles
front-loaded weights; color identifies the testing method.}
\label{fig:sim-weights}
\end{figure}

Figure~\ref{fig:sim-weights} illustrates sensitivity to
prespecified allocations. Front-loaded weights can help when
signals are concentrated early and hurt when they occur late.
Consequently, the choice of weights should be connected to
external ordering information rather than tuned to the observed
$p$-values. 

\paragraph{Interpretation and limitations.}
The experiments do not establish a universal power advantage
for the proposed methods. In the representative independence
comparison, the position-dependent calibration improves on
Lynch's $K=m$ rule and LOND, but LORD++ has higher power.
Uncorrected fallback can improve on conventional fallback;
the size of this benefit depends on signal placement and
allocation weights. The corrected fallback rule has lower
estimated power than conventional fallback in all 20 independent
core scenarios. Since conventional fallback already controls
the FWER under arbitrary dependence, the correction supplies
validity for the enlarged fallback construction rather than
broader protection relative to that comparator.
The PLTDN experiment illustrates a separate route to
calibration for two-sided tests with known covariance.

These conclusions are limited to the specified Gaussian
models and allocation schedules. In particular, the AR(1)
grid does not represent all joint distributions covered by
an arbitrary-dependence theorem. Each random fixed arrangement
uses a single prespecified configuration, so the study does
not average over all possible signal placements. The
experiments also do not assess covariance estimation or
optimize the allocation schedule of any method. Numerical
results complement, rather than establish, the theoretical
FDR guarantees.

\section{Discussion and Conclusions}
\label{s:discussion}

We have studied FDR control for a fixed family of
hypotheses tested sequentially in a prespecified order.
Decisions use only currently available $p$-values,
are not subsequently revised, and permit continued
testing after non-rejections. Our development distinguishes
two threshold structures. Generalized fixed-sequence
procedures use the testing position and the number of
preceding discoveries. Fallback-type procedures
additionally use recursively transferred testing levels,
distinguishing histories with the same rejection count
but different arrangements of rejections.

For rejection-count procedures, we studied a
position-dependent calibration under independence and
a linear rule calibrated through null marginal
distribution functions under PLTDN. These results are
closely connected to existing sequential methods.
In particular, the basic linear threshold form coincides
with LOND, while the position-dependent procedure is
related to the generalized fixed-sequence procedures
of \citet{lynch2017control}. The developments here
concern the specific calibrations and dependence
conditions, rather than rejection-count adaptation
itself.

The fallback construction combines two features of
the testing history: the total number of preceding
discoveries and accumulated allocations along consecutive
rejections. We established FDR control under
history-based PRDS and, with a position-specific
correction, under arbitrary dependence. This construction
connects conventional fallback testing with sequential
FDR control while allowing later hypotheses to remain
testable after non-rejections.

The simulations identify both benefits and limitations.
The position-dependent procedure improved on the
non-stopping Lynch benchmark and LOND in the independent
core scenarios, although the gains were modest.
PRDS fallback improved on conventional fallback in
those scenarios, with gains depending on signal placement.
Neither construction uniformly outperformed established
online methods. The arbitrary-dependence correction
was conservative and could reduce power below that of
conventional fallback, which already controls the FWER
and hence the FDR under arbitrary dependence.
Thus, its validity does not by itself imply a practical
advantage over that benchmark. The two-sided experiments
also supported the feasibility of PLTDN calibration
when the dependence parameters are known.

These findings motivate further work on calibration.
A central question is whether the arbitrary-dependence
fallback correction can be sharpened to preserve more
of the benefit from rejection-count adaptation.
Another is how external information can guide allocation
weights without compromising validity. For PLTDN-based
procedures, replacing known dependence parameters with
estimates requires additional analysis. Broader numerical
studies could assess these choices beyond the Gaussian
models and fixed signal configurations considered here.

Corollary~\ref{cor:prefix-fdr} extends the terminal results
to every deterministic testing time within the planned
horizon, with that horizon retained in the thresholds.
This provides pointwise-in-time FDR control but does not
establish validity at arbitrary data-dependent stopping
times or across repeated monitoring periods. Unknown
horizons, groups of hypotheses arriving at each testing
occasion, and more general level-transfer mechanisms
require separate treatment. The distinction between
rejection-count adaptation and recursive level transfer
provides a useful way to formulate these extensions
and identify the assumptions needed for their FDR
control.

\clearpage
\begin{appendices}

\section{Auxiliary probability bounds and a sequential FDR bound}
\label{app:probability-bounds}

We use the notation of Section~\ref{s:setup}.
All empty sums are zero. The zero-threshold convention applies
throughout. Marginal validity is used except in the PLTDN proof,
where it is replaced by the stated marginal calibration.

The next two probability bounds are discrete specializations of
established dependency-control inequalities; see
\citet[Lemma~3.2(ii)--(iii)]{blanchard2008two}.
They are stated with proofs to make the subsequent arguments
self-contained. No independence assumption is required.

\begin{lemma}[A bound under arbitrary dependence]
\label{lem:seq-arbitrary}
Let $U\in[0,1]$ satisfy $\Pr(U\le u)\le u$ for $u\in[0,1]$,
and let $K\in\{1,\ldots,M\}$ be an arbitrary random variable.
For deterministic numbers $0=s_0\le s_1\le\cdots\le s_M$,
\[
\mathbb{E}\!\left[\frac{\mathbf{1}\{U\le s_K\}}{K}\right]
\le\sum_{k=1}^{M}\frac{s_k-s_{k-1}}{k}.
\]
The numbers $s_k$ need not be bounded by one.
\end{lemma}
\begin{proof}
Monotonicity of the thresholds gives the pointwise inequality
\[
\frac{\mathbf{1}\{U\le s_K\}}{K}
\le\sum_{k=1}^{M-1}\left(\frac1k-\frac1{k+1}\right)
\mathbf{1}\{U\le s_k\}
+\frac1M\mathbf{1}\{U\le s_M\}.
\]
Taking expectations, using $\Pr(U\le s_k)\le s_k$, and summing
by parts proves the assertion. The probability bound also holds for
$s_k>1$, since probabilities do not exceed one.
\end{proof}

\begin{lemma}[A bound under positive dependence]
\label{lem:seq-prds}
Let $U$ be marginally valid and let $K\in\{1,\ldots,M\}$.
Suppose that, for each $k$, the function
\[
u\longmapsto\Pr(K\ge k\mid U\le u)
\]
is nonincreasing wherever the conditioning event has positive
probability. Then, for every $c\ge0$,
\[
\mathbb{E}\!\left[\frac{\mathbf{1}\{U\le cK\}}{K}\right]\le c.
\]
\end{lemma}
\begin{proof}
For $c=0$, the assertion follows from $\Pr(U=0)=0$.
Otherwise, put $s_k=\min\{1,ck\}$ and let $k_0$ be the smallest
index with $\Pr(U\le s_{k_0})>0$. If no such index exists, the
expectation is zero. Marginal validity implies
\[
\mathbb{E}\!\left[\frac{\mathbf{1}\{U\le cK\}}{K}\right]
\le c\sum_{k=k_0}^{M}\Pr(K=k\mid U\le s_k).
\]
The sum on the right equals
\begin{align*}
&\Pr(K\ge k_0\mid U\le s_{k_0})\\
&\quad+\sum_{k=k_0+1}^{M}
\left\{\Pr(K\ge k\mid U\le s_k)
-\Pr(K\ge k\mid U\le s_{k-1})\right\}
\le1.
\end{align*}
This proves the result.
\end{proof}

\subsection{A supporting sequential FDR bound}
\label{app:proof-fdr}

The following is the sequential denominator argument underlying
LOND-type bounds \citep{javanmard2015online}.
Its first assertion requires neither marginal validity nor a
dependence assumption.

\begin{lemma}[History-based FDR bound]
\label{FDR}
For predictable $T_i\in[0,1]$, define
\[
B_i=\mathbb E\!\left[
\frac{\Pr(P_i\le T_i\mid\mathcal F_{i-1})}{R_{i-1}+1}
\right],\qquad i\in I_0.
\]
Then $\mathrm{FDR}\le\sum_{i\in I_0}B_i$.
For deterministic thresholds
$T_i=\widetilde\alpha_i(R_{i-1})$, put
\[
q_i(r)=\Pr\{P_i\le\widetilde\alpha_i(r)\mid\mathcal F_{i-1}\}.
\]
Then
\begin{align*}
B_i
&=\Pr\{P_i\le\widetilde\alpha_i(0)\}\\
&\quad+\sum_{r=1}^{i-1}\mathbb E\!\left[
\left\{\frac{q_i(r)}{r+1}-\frac{q_i(r-1)}r\right\}
\mathbf1\{R_{i-1}\ge r\}\right].
\end{align*}
The sum is empty for $i=1$.
\end{lemma}
\begin{proof}
On the event $\{D_i=1\}$, we have
$R_m\ge R_{i-1}+1$. Consequently,
\[
\frac{D_i}{R_m\vee1}
\le
\frac{D_i}{R_{i-1}+1}.
\]
Since $D_i\le\mathbf{1}\{P_i\le T_i\}$, including under the
zero-threshold convention, summing over $i\in I_0$,
taking expectations, and conditioning on
$\mathcal{F}_{i-1}$ gives
\[
\mathrm{FDR}
=
\sum_{i\in I_0}
\mathbb{E}\!\left[
\frac{D_i}{R_m\vee1}
\right]
\le
\sum_{i\in I_0}B_i.
\]

For the deterministic rejection-count rule, set
\[
q_i(r)
=
\Pr\!\left(
P_i\le\widetilde{\alpha}_i(r)
\mid\mathcal{F}_{i-1}
\right).
\]
The finite telescoping identity
\[
\frac{q_i(R_{i-1})}{R_{i-1}+1}
=
q_i(0)
+
\sum_{r=1}^{i-1}
\left\{
\frac{q_i(r)}{r+1}
-
\frac{q_i(r-1)}{r}
\right\}
\mathbf{1}\{R_{i-1}\ge r\}
\]
holds pointwise. Taking expectations and using
\[
\mathbb{E}[q_i(0)]
=
\Pr\!\left(P_i\le\widetilde{\alpha}_i(0)\right)
\]
establishes the stated representation.
\end{proof}

\section{Proof of the position-dependent independence result}
\label{APP:A6}

We prove Theorem~\ref{thm:position-indep} for
Procedure~\ref{proc:gfs-indep}. Throughout this section,
the $p$-values are mutually independent and true-null
$p$-values are marginally valid. If $m=1$, the threshold
is $\alpha$, so the conclusion follows immediately.
Assume henceforth that $m\ge2$.
The rejection-index decomposition follows the approach of
\citet{lynch2017control}.

\subsection{Notation indexed by successive rejections}
\label{app:rejection-notation}

Let $J_k$ be the position of the $k$th rejection, with $J_0=0$,
and set $J_k=m+1$ if fewer than $k$ rejections occur.
Let $V_k^\star$ and $S_k^\star$ count false and true discoveries
among the first $\min\{k,R_m\}$ rejections. These quantities are
distinct from $V_i$ and $S_i$, which count discoveries among the
first $i$ hypotheses. Define
\[
Q_0=0,\qquad G_0=1,\qquad
Q_k=\frac{V_k^\star}{k},\quad
G_k=\frac{S_k^\star}{k},\quad k=1,\ldots,m.
\]
The value $G_0=1$ is an auxiliary convention, not a definition of
$S_0^\star/0$. On $\{J_k\le m\}$, $Q_k+G_k=1$.
For $k=1,\ldots,m$ and $j=0,\ldots,m$, put
\[
W_k(j)=\mathbf{1}\{J_{k-1}\le j<J_k\}.
\]
For $0\le k<m$ and $j=0,\ldots,m$, define
\[
f_k(j)=\frac{\alpha(m-j)}{m-k}G_k,
\qquad f_m(j)=0.
\]
Also set $f_k(m+1)=0$ for every $k$. In particular,
$f_0(0)=\alpha$ and $f_k(m)=0$.

The following pathwise identities hold:
\begin{align}
\mathbf{1}\{J_k=j\}
&=W_k(j-1)\mathbf{1}\{P_j\le\gamma_{kj}\},
\label{eq:gfs-Ji-identity}\\
W_k(j)-W_k(j-1)
&=\mathbf{1}\{J_{k-1}=j\}-\mathbf{1}\{J_k=j\},
\label{eq:gfs-Wi-difference}
\end{align}
where $1\le k\le j\le m$ and
\[
\gamma_{kj}
=\frac{k\alpha}{d_{kj}},\qquad
 d_{kj}=j\alpha+(m-k+1)(1-\alpha).
\]
On $\{W_k(j-1)=1\}$, exactly $k-1$ rejections precede $H_j$,
so $\gamma_{kj}$ is its testing threshold.

\begin{lemma}[Rejection-index decomposition]
\label{lem:gfs-indep-fdr}
For any sequential procedure,
\[
\mathrm{FDR}
=\sum_{k=1}^{m}\mathbb{E}\!\left[
(Q_k-Q_{k-1})\mathbf{1}\{J_k\le m\}\right].
\]
\end{lemma}
\begin{proof}
On $\{R_m=r\ge1\}$, the sum inside the expectation telescopes
to $Q_r=V_m/r$. On $\{R_m=0\}$, it equals zero.
\end{proof}

\begin{lemma}[Adjusted rejection increment]
\label{lem:gfs-indep-aux1}
For $1\le k\le j\le m$,
\begin{align*}
&\mathbb{E}\!\left[
\{Q_k-Q_{k-1}+f_k(j)-f_{k-1}(j)\}
\mathbf{1}\{J_k=j\}\right]\\
&\hspace{15mm}\le
\frac{\alpha}{m-k+1}\,
\mathbb{E}[G_{k-1}W_k(j-1)].
\end{align*}
\end{lemma}
\begin{proof}
All calculations below concern the event $W_k(j-1)=1$.
Restricted to this event, the preceding $k-1$ discoveries and their
types are determined by $\mathcal F_{j-1}$. More formally, their
products with $W_k(j-1)$ have $\mathcal F_{j-1}$-measurable versions.
For $j\in I_0$, independence gives
\[
\Pr(P_j\le\gamma_{kj}\mid\mathcal F_{j-1})
=\Pr(P_j\le\gamma_{kj})\le\gamma_{kj}.
\]
We use this inequality below only with nonnegative, history-measurable
coefficients.

\emph{First rejection: $k=1$.}
If $j\in I_0$, the adjusted increment on $\{J_1=j\}$ is
\[
1-\frac{\alpha(m-j)}m=\frac{d_{1j}}m.
\]
It is positive, and its product with $\gamma_{1j}$ is
$\alpha/m$. Independence and marginal validity therefore give the
required bound. If $j\in I_1$, the adjusted increment equals
\[
\frac{\alpha(m-j)}{m(m-1)}\le\frac\alpha m.
\]
Multiplying by $\mathbf{1}\{J_1=j\}\le W_1(j-1)$ proves the claim.

\emph{Intermediate rejections: $2\le k<m$.}
Write $v=V_{k-1}^\star$ and $s=S_{k-1}^\star$ on the
preceding-history event, so $v+s=k-1$.
If $j\in I_0$, the adjusted increment on $\{J_k=j\}$ is
\[
A_{kj}^{(0)}=
\frac{\{(m-k)(m-k+1)+\alpha(m-j)(2k-m-1)\}s}
{k(k-1)(m-k)(m-k+1)}.
\]
This quantity is nonnegative: when $2k-m-1<0$, use
$m-j\le m-k$ to bound the expression in braces from below
by $(m-k)\{m-k+1+\alpha(2k-m-1)\}>0$;
otherwise nonnegativity is immediate.
A direct calculation gives
\[
\frac{\alpha s}{(m-k+1)(k-1)}
-A_{kj}^{(0)}\gamma_{kj}
=
\frac{\alpha^2(j-k)s}{(m-k)(m-k+1)d_{kj}}\ge0.
\]
The conditional probability bound above proves the desired inequality.

If $j\in I_1$, the adjusted increment is
\[
A_{kj}^{(1)}=
-\frac{v}{k(k-1)}
+\alpha(m-j)\left\{
\frac{s+1}{k(m-k)}-\frac{s}{(k-1)(m-k+1)}
\right\}.
\]
The coefficient of $\alpha(m-j)$ is nonnegative for
$0\le s\le k-1$. Using $m-j\le m-k$ therefore yields
\[
A_{kj}^{(1)}
\le\frac{\alpha s}{(m-k+1)(k-1)}
-\frac{(1-\alpha)v}{k(k-1)}
\le\frac{\alpha s}{(m-k+1)(k-1)}.
\]
The right-hand side is nonnegative. Multiplication by
$\mathbf{1}\{J_k=j\}\le W_k(j-1)$ proves the bound.

\emph{Final rejection: $k=m$.}
Only $j=m$ is possible. Both potential terms at $j=m$ vanish.
For $j\in I_0$, the adjusted increment is $G_{m-1}/m$,
and
\[
\frac{\gamma_{mm}}m
=\frac{\alpha}{m\alpha+1-\alpha}\le\alpha.
\]
For $j\in I_1$, the increment is nonpositive. These observations
complete the proof.
\end{proof}

\begin{lemma}[Potential-function identity]
\label{lem:gfs-indep-aux2}
For $k=1,\ldots,m$,
\begin{align*}
&\mathbb{E}\!\left[
f_{k-1}(J_{k-1})\mathbf{1}\{J_{k-1}\le m\}
-f_{k-1}(J_k)\mathbf{1}\{J_k\le m\}\right]\\
&\hspace{15mm}=
\frac{\alpha}{m-k+1}\sum_{j=k}^{m}
\mathbb{E}[G_{k-1}W_k(j-1)].
\end{align*}
\end{lemma}
\begin{proof}
If the $(k-1)$st rejection does not occur, both sides vanish
pathwise. Otherwise, $W_k(j-1)=1$ precisely between that rejection
and the next rejection, or the end of the sequence if there is no
next rejection. Over this interval,
\[
f_{k-1}(j-1)-f_{k-1}(j)
=\frac{\alpha}{m-k+1}G_{k-1}.
\]
Summing these differences, using $f_{k-1}(m)=0$, proves the
identity pathwise and hence in expectation. The same calculation
applies to $k=1$, with $J_0=0$ and $G_0=1$.
\end{proof}

\subsection{Completion of the proof of
Theorem~\ref{thm:position-indep}}
\begin{proof}
Let
\begin{align*}
\mathcal A_k&=\sum_{j=k}^{m}\mathbb{E}\!\left[
\{Q_k-Q_{k-1}+f_k(j)-f_{k-1}(j)\}
\mathbf{1}\{J_k=j\}\right],\\
\mathcal C_k&=\mathbb{E}\!\left[
f_{k-1}(J_{k-1})\mathbf{1}\{J_{k-1}\le m\}
-f_{k-1}(J_k)\mathbf{1}\{J_k\le m\}\right].
\end{align*}
Lemmas~\ref{lem:gfs-indep-aux1} and~\ref{lem:gfs-indep-aux2}
give $\mathcal A_k\le\mathcal C_k$.
By Lemma~\ref{lem:gfs-indep-fdr},
\begin{align*}
\mathrm{FDR}
&=\sum_{k=1}^{m}\mathcal A_k
-\sum_{k=1}^{m}\mathbb{E}\!\left[
\{f_k(J_k)-f_{k-1}(J_k)\}\mathbf{1}\{J_k\le m\}\right]\\
&\le\sum_{k=1}^{m}\mathbb{E}\!\left[
f_{k-1}(J_{k-1})\mathbf{1}\{J_{k-1}\le m\}
-f_k(J_k)\mathbf{1}\{J_k\le m\}\right]\\
&=f_0(0)=\alpha,
\end{align*}
where the final equality uses telescoping and $f_m\equiv0$.
\end{proof}

\section{Proof of the sequential PLTDN calibration}
\label{app:pltdn-proof}
\label{APP:A1}
\label{app:gfs-proofs}
\subsection{Proof of Theorem~\ref{thm:FDR_PLTDN}}
\label{app:proof-pltdn}

We analyze Procedure~\ref{proc:gfs-pltdn}.
The functions $F_i$ are deterministic marginal upper bounds;
super-uniformity is not assumed in this proof.
The proof applies the PLTDN argument of
\citet{GhoshSarkar2025} to the preceding rejection count.

\begin{proof}
Fix $i\in I_0$ with $a_i>0$, and set
\[
t_r=\min\{1,(r+1)a_i\},\qquad r=0,\ldots,i-1.
\]
The linear thresholds are nondecreasing in the preceding
rejection count. Induction therefore shows that $R_{i-1}$
is coordinatewise nonincreasing in $\boldsymbol P_{<i}$,
including under the zero-threshold convention. Thus
$A_r=\{R_{i-1}\ge r\}$ is a decreasing history event. It can be written as
$\{h(\boldsymbol{P}_{<i})\le-r\}$ with
$h=-R_{i-1}$ coordinatewise nondecreasing.
If $\Pr(A_r)>0$, history-based PLTDN implies
\[
\frac{\Pr(P_i\le t_r\mid A_r)}{r+1}
\le\frac{\Pr(P_i\le t_{r-1}\mid A_r)}{r},
\qquad r\ge1.
\]
Indeed, $t_r\ge t_{r-1}>0$ and
\[
\frac{\Pr(P_i\le t_r\mid A_r)}{r+1}
=\frac{\Pr(P_i\le t_r\mid A_r)}{t_r}
  \frac{t_r}{r+1}
\le\frac{\Pr(P_i\le t_{r-1}\mid A_r)}{t_{r-1}}
  \frac{t_{r-1}}r.
\]
Here PLTDN bounds the first factor and truncation gives
$t_r/(r+1)\le t_{r-1}/r$. Events $A_r$ of probability
zero contribute zero and require no conditional probability.
Thus every expected increment in the telescoping representation
of $B_i$ in Lemma~\ref{FDR} is nonpositive. Consequently,
\[
B_i\le\Pr(P_i\le a_i)\le F_i(a_i).
\]
If $a_i=0$, the definition of $B_i$ in Lemma~\ref{FDR}
gives
\[
B_i=\mathbb E\!\left[
\frac{\Pr(P_i=0\mid\mathcal F_{i-1})}{R_{i-1}+1}
\right]\le\Pr(P_i=0)\le F_i(0).
\]
Thus the same bound holds, whether zero-threshold decisions are
excluded by convention or included in the upper bound $B_i$. Summing yields
\[
\mathrm{FDR}\le\sum_{i\in I_0}F_i(a_i)
\le\sum_{i=1}^{m}F_i(a_i)\le\alpha.
\]
\end{proof}

\subsection{Gaussian verification of history-based PLTDN}
\label{app:gaussian-pltdn}

The following lemma verifies the transformed-testing
construction used in Section~\ref{s:sim}, within the
dependence-adjusted framework of
\citet{Sarkar_2025,GhoshSarkar2025}.
Write $\overline F_0$ for the survival function of
a central $\chi^2_1$ variable and $q(t)=\overline F_0^{-1}(t)$.

\begin{lemma}[Gaussian history-based PLTDN]
\label{lem:gaussian-history-pltdn}
Let $\boldsymbol X\sim\mathcal N_m(\boldsymbol\mu,\Sigma)$,
where $\Sigma$ is positive definite with unit diagonal.
For $i\ge2$, put
\[
\tau_i=\operatorname{Var}(X_i\mid X_1,\ldots,X_{i-1}),
\qquad \tau_1=1,
\]
so $0<\tau_i\le1$. Define
\[
\widetilde P_i=\overline F_0(X_i^2/\tau_i),
\qquad
F_i(t)=\overline F_0\bigl(\tau_i q(t)\bigr),
\quad 0<t<1,
\]
with $F_i(0)=0$ and $F_i(1)=1$.
For every true null $H_i:\mu_i=0$,
\[
\Pr(\widetilde P_i\le t)=F_i(t),
\qquad 0\le t\le1.
\]
Moreover, the ordered quantities
$(\widetilde P_1,\ldots,\widetilde P_m)$ satisfy
history-based PLTDN in Definition~\ref{def:history-pltdn}.
\end{lemma}

\begin{proof}
Fix $i\in I_0$ and let
$\mathcal G_{i-1}=\sigma(X_1,\ldots,X_{i-1})$.
Gaussian conditioning gives
\[
X_i\mid\mathcal G_{i-1}
\sim\mathcal N(\nu_i,\tau_i),
\qquad
\nu_i=\mathbb E[X_i\mid\mathcal G_{i-1}].
\]
For $i=1$, use the trivial history and $\nu_1=0$.
Consequently, $X_i^2/\tau_i$ has conditional
noncentral $\chi^2_1$ distribution with
$\lambda_i=\nu_i^2/\tau_i\ge0$.

Let $f_\lambda$ and $\overline F_\lambda$ denote
the density and survival function of a noncentral
$\chi^2_1$ variable with parameter $\lambda$.
For $x>0$,
\[
\frac{f_\lambda(x)}{f_0(x)}
=\exp(-\lambda/2)\cosh(\sqrt{\lambda x}),
\]
which is nondecreasing in $x$. Therefore
\[
\frac{\overline F_\lambda(x)}{\overline F_0(x)}
=\mathbb E_0\!\left[
\frac{f_\lambda(Y)}{f_0(Y)}\,\middle|\,Y\ge x
\right]
\]
is nondecreasing in $x$, where $Y\sim\chi^2_1$
under $\mathbb E_0$.
Since $q(t)$ decreases with $t$,
\[
\frac{\Pr(\widetilde P_i\le t\mid\mathcal G_{i-1})}{t}
=\frac{\overline F_{\lambda_i}(q(t))}
        {\overline F_0(q(t))}
\]
is nonincreasing in $t\in(0,1)$, with the
corresponding endpoint value at $t=1$ obtained
by continuity.

The preceding transformed quantities are measurable
with respect to $\mathcal G_{i-1}$. Hence, for any
event $A$ determined by them with $\Pr(A)>0$,
\[
\frac{\Pr(\widetilde P_i\le t\mid A)}{t}
=\mathbb E\!\left[
\frac{\Pr(\widetilde P_i\le t\mid\mathcal G_{i-1})}{t}
\,\middle|\, A\right]
\]
is also nonincreasing in $t$. In particular, this
holds for every decreasing history event required
by Definition~\ref{def:history-pltdn}.

Finally, under $\mu_i=0$, the marginal distribution
of $X_i^2$ is central $\chi^2_1$. Thus
\[
\Pr(\widetilde P_i\le t)
=\Pr\{X_i^2\ge\tau_i q(t)\}
=\overline F_0\bigl(\tau_i q(t)\bigr)=F_i(t).
\]
The endpoint identities follow by continuity.
\end{proof}

For the AR(1) covariance used in the simulations,
$\tau_1=1$ and $\tau_i=1-\rho^2$ for $i\ge2$.
The lemma therefore applies for both signs of $\rho$,
provided $|\rho|<1$. The transformation and its
marginal calibration use the known covariance;
the lemma does not justify substituting estimated
conditional variances.

\section{Proofs for the fallback procedures}
\label{app:fallback-proofs}

\subsection{A common first-false-rejection bound}
\label{app:first-false-bound}

The two fallback procedures are analyzed separately, each with
its own rejection indicators, counts, and transferred levels.
The decomposition below follows the first-false-rejection
approach used by \citet{lynch2017control}.

If $I_0$ is empty, the FDR is zero. Otherwise, write its ordered
indices as $u_1<\cdots<u_{m_0}$ and put $u_0=0$.
Let $U$ be the index of the first rejected true null, with
$U=m+1$ if none is rejected. For $\ell=1,\ldots,m_0$, set
\[
q_\ell=m-u_\ell+1,\qquad
b_\ell=\alpha\sum_{j=u_{\ell-1}+1}^{u_\ell}\omega_j,
\qquad K_\ell=q_\ell+R_{u_\ell-1}.
\]
On $\{U=u_\ell\}$, every preceding rejection is a true discovery,
so $S_m\ge R_{u_\ell-1}$, while $V_m\le q_\ell$. Since $v/(v+s)$ increases with $v$ and decreases with $s$,
\[
\frac{V_m}{R_m\vee1}\mathbf{1}\{U=u_\ell\}
\le\frac{q_\ell}{K_\ell}\mathbf{1}\{U=u_\ell\}.
\]
Moreover, if $\ell>1$, the preceding true null at $u_{\ell-1}$
was not rejected and its carried-forward level was not transferred.
For $\ell=1$, accumulation starts at the beginning of the sequence.
In either case, the recursion gives
\[
L_{u_\ell}\le b_\ell\quad\hbox{on }\{U=u_\ell\}.
\]
If $b_\ell=0$, this event is empty by the zero-level convention.
For $b_\ell>0$, the PRDS and corrected rules therefore give,
respectively,
\begin{align}
\mathbb E\!\left[
\frac{V_m}{R_m\vee1}\mathbf1\{U=u_\ell\}\right]
&\le q_\ell\mathbb E\!\left[
\frac{\mathbf1\{P_{u_\ell}\le K_\ell b_\ell/q_\ell\}}
{K_\ell}\right],\label{eq:fb-key-bound-prds}\\
\mathbb E\!\left[
\frac{V_m}{R_m\vee1}\mathbf1\{U=u_\ell\}\right]
&\le q_\ell\mathbb E\!\left[
\frac{\mathbf1\{P_{u_\ell}\le K_\ell b_\ell/(q_\ell c_{u_\ell})\}}
{K_\ell}\right].\label{eq:fb-key-bound-arb}
\end{align}
To obtain these bounds, use the threshold implication on
$\{U=u_\ell\}$ and then drop the indicator of this event
from the nonnegative upper bound. Threshold truncation at one
does not affect the implication, because $P_{u_\ell}\le1$.
Each inequality concerns the history generated by its own procedure.

\subsection{Proof of Theorem~\ref{thm:fb-prds}}
\label{APP:A7}
\begin{proof}
The recursive levels $L_i$ are nondecreasing functions of preceding
rejection indicators, as is the multiplier
$1+R_{i-1}/(m-i+1)$. Induction on $i$ therefore shows that
$R_i$ is coordinatewise nonincreasing in $(P_1,\ldots,P_i)$.
The zero-level convention preserves this monotonicity.

Consequently, $\{K_\ell\ge k\}$ is a decreasing event in
$\boldsymbol P_{<u_\ell}$. History-based PRDS supplies the
conditional monotonicity required by Lemma~\ref{lem:seq-prds}.
Applying that lemma to \eqref{eq:fb-key-bound-prds}, with
$c=b_\ell/q_\ell$, bounds the $\ell$th contribution by $b_\ell$.
Thus
\[
\mathrm{FDR}\le\sum_{\ell=1}^{m_0}b_\ell
=\alpha\sum_{j=1}^{u_{m_0}}\omega_j\le\alpha.
\]
\end{proof}

\subsection{Proof of Theorem~\ref{thm:fb-arb}}
\label{APP:A8}
\begin{proof}
Fix $\ell$ with $b_\ell>0$. Since
$q_\ell\le K_\ell\le m$, define the deterministic sequence
\[
s_k=\begin{cases}
0,&0\le k<q_\ell,\\[2pt]
\displaystyle\frac{k b_\ell}{q_\ell c_{u_\ell}},
&q_\ell\le k\le m.
\end{cases}
\]
Lemma~\ref{lem:seq-arbitrary} gives
\begin{align*}
\mathbb E\!\left[
\frac{\mathbf1\{P_{u_\ell}\le s_{K_\ell}\}}{K_\ell}\right]
&\le\sum_{k=1}^{m}\frac{s_k-s_{k-1}}k\\
&=\frac{b_\ell}{q_\ell c_{u_\ell}}
\left(1+\sum_{k=q_\ell+1}^{m}\frac1k\right)
=\frac{b_\ell}{q_\ell}.
\end{align*}
The first nonzero threshold increment occurs at $k=q_\ell$;
its contribution is $b_\ell/(q_\ell c_{u_\ell})$.
The final equality follows from
$c_{u_\ell}=1+H_m-H_{q_\ell}$.
Combining this bound with \eqref{eq:fb-key-bound-arb} and summing
as in the PRDS proof yields $\mathrm{FDR}\le\alpha$.
\end{proof}

\section{Verification of established LOND calibrations}
\label{app:arbitrary-proofs}

These arguments verify the PRDS and arbitrary-dependence
calibrations stated in the remarks of Section~\ref{s:Depend}.
They are included for completeness, rather than as additional
proposed procedures.

\subsection{PRDS calibration in Remark~\ref{thm:FDR_PRDS}}
\label{app:proof-prds}
\begin{proof}
The thresholds are nondecreasing in the number of preceding
rejections. Induction on the testing position therefore shows that
$R_{i-1}$ is coordinatewise nonincreasing in
$\boldsymbol{P}_{<i}$. In particular, each event
$\{R_{i-1}+1\ge k\}$ is decreasing in that vector.
History-based PRDS implies that its conditional probability given
$P_i\le u$ is nonincreasing in $u$.

Apply Lemma~\ref{lem:seq-prds} with $U=P_i$,
$K=R_{i-1}+1$, and $c=a_i$. Lemma~\ref{FDR} then gives
\[
\mathrm{FDR}
\le\sum_{i\in I_0}
\mathbb{E}\!\left[\frac{\mathbf{1}\{P_i\le (R_{i-1}+1)a_i\}}
{R_{i-1}+1}\right]
\le\sum_{i\in I_0}a_i\le\alpha.
\]
Truncating thresholds at one does not change their rejection events.
\end{proof}

\subsection{Arbitrary-dependence calibration in
Remark~\ref{thm:FDRArbDep}}
\label{app:proof-harmonic}

\begin{proof}
For each $i\in I_0$, put $K_i=R_{i-1}+1\in\{1,\ldots,i\}$
and $s_k=ka_i/H_i$ for $k=0,\ldots,i$.
Lemma~\ref{lem:seq-arbitrary} gives
\[
\mathbb E\!\left[
\frac{\mathbf1\{P_i\le K_i a_i/H_i\}}{K_i}
\right]
\le\sum_{k=1}^i\frac{s_k-s_{k-1}}k
=\frac{a_i}{H_i}\sum_{k=1}^i\frac1k=a_i.
\]
The lemma allows $s_k>1$, so threshold truncation poses no
problem. Combining this bound with Lemma~\ref{FDR} yields
\[
\mathrm{FDR}\le\sum_{i\in I_0}a_i\le\alpha.
\]
This recovers the dependence-adjusted LOND guarantee of
\citet[Theorem~2.7]{javanmard2015online}.
\end{proof}

\section{FDR Control at Deterministic Testing Times}
\label{app:prefix-fdr}

\begin{proof}[Proof of Corollary~\ref{cor:prefix-fdr}]
Fix $t\in\{1,\ldots,m\}$. All thresholds and
allocations retain their original specifications
for the planned horizon $m$.

For Procedure~\ref{proc:gfs-pltdn}, apply the
sequential denominator argument of Lemma~\ref{FDR}
to the first $t$ tests. The proof of
Theorem~\ref{thm:FDR_PLTDN} bounds each true-null
contribution by $F_i(a_i)$, independently of
whether tests after position $t$ are performed.
Consequently,
\[
\mathrm{FDR}(t)
\le\sum_{\substack{i\le t\\i\in I_0}}F_i(a_i)
\le\sum_{i=1}^m F_i(a_i)\le\alpha.
\]

For Procedures~\ref{proc:gfs-indep},
\ref{proc:fb-prds}, and~\ref{proc:fb-arb},
construct a modified family that preserves the
hypotheses and joint distribution of
$(P_1,\ldots,P_t)$ and replaces every later
hypothesis by a true null with $P_i'=1$ for $i>t$.
Predictability and the unchanged horizon imply
that all decisions through position $t$ agree
with the original procedure.

The added constant $p$-values are marginally valid,
and their addition preserves mutual independence
when required. The history-based PRDS conditions
for true nulls among the first $t$ tests are
unchanged. For an added true null, the event
$\{P_i'\le u\}$ has positive probability only
at $u=1$, so its PRDS condition is satisfied.
Thus the corresponding terminal theorem applies
to the modified family under each of the three
sets of assumptions.

Let $A$ denote the number of rejections after
position $t$ in the modified family. Every such
rejection is false, giving
\[
V_m'=V_t+A,
\qquad R_m'=R_t+A.
\]
Since $0\le V_t\le R_t$ and $A\ge0$,
\[
\frac{V_t}{R_t\vee1}
\le\frac{V_t+A}{(R_t+A)\vee1}
=\frac{V_m'}{R_m'\vee1}
\]
pathwise. Taking expectations and applying the
appropriate terminal theorem proves
$\mathrm{FDR}(t)\le\alpha$. As $t$ was arbitrary,
the conclusion holds at every deterministic time
within the planned horizon.
\end{proof}

\end{appendices}

\clearpage
\bibliographystyle{plainnat}
\bibliography{ref}

\end{document}